# Heat shock partially dissociates the overlapping modules of the yeast protein-protein interaction network: a systems level model of adaptation

Ágoston Mihalik and Peter Csermely[1]

Department of Medical Chemistry, Semmelweis University, P O Box 260., H-1444 Budapest 8, Hungary

**ABSTRACT**

Network analysis became a powerful tool giving new insights to the understanding of cellular behavior. Heat shock, the archetype of stress responses, is a well-characterized and simple model of cellular dynamics. *S. cerevisiae* is an appropriate model organism, since both its protein-protein interaction network (interactome) and stress response at the gene expression level have been well characterized. However, the analysis of the reorganization of the yeast interactome during stress has not been investigated yet. We calculated the changes of the interaction-weights of the yeast interactome from the changes of mRNA expression levels upon heat shock. The major finding of our study is that heat shock induced a significant decrease in both the overlaps and connections of yeast interactome modules. In agreement with this the weighted diameter of the yeast interactome had a 4.9-fold increase in heat shock. Several key proteins of the heat shock response became centers of heat shock-induced local communities, as well as bridges providing a residual connection of modules after heat shock. The observed changes resemble to a 'stratus-cumulus' type transition of the interactome structure, since the unstressed yeast interactome had a globally connected organization, similar to that of stratus clouds, whereas the heat shocked interactome had a multifocal organization, similar to that of cumulus clouds. Our results showed that heat shock induces a partial disintegration of the global organization of the yeast interactome. This change may be rather general occurring in many types of stresses. Moreover, other complex systems, such as single proteins, social networks and ecosystems may also decrease their inter-modular links, thus develop more compact modules, and display a partial disintegration of their global structure in the initial phase of crisis. Thus, our work may provide a model of a general, system-level adaptation mechanism to environmental changes.

**AUTHOR SUMMARY** In the last two decades our knowledge on stress-induced changes has been expanded rapidly. As a part of this work a large number of key proteins and biological processes of cellular adaptation to stress have been uncovered. However, we know relatively little on the systems level changes of the cell in stress. In our study we used the network approach to study the changes of the yeast protein-protein interaction network (interactome) in the archetype of stress, heat shock. The major finding of our study is that heat shock induced a marked decrease in the inter-community connections of the yeast interactome. The observed changes resembled to a 'stratus-cumulus' type transition of the interactome structure, since the unstressed yeast interactome had a globally connected organization, similar to that of stratus clouds, whereas the heat shocked interactome had a multifocal organization, similar to that of cumulus clouds. Our results indicated that heat shock induces a partial disintegration of the global protein-protein network structure of yeast cells. This change may be rather general occurring at the initial phase of crises in many complex systems, such as proteins in physical stretch, ecosystems in abrupt environmental changes or social networks in economic crisis.

[a]*Corresponding author:* Csermely, P. (csermely@eok.sote.hu)

# INTRODUCTION

In the last decade due to the advance of high-throughput technologies system level inquiries became widespread. The network approach emerged as a versatile tool to assess the background of the regulation and changes of cellular functions. Analysis of protein-protein interaction (PPI) networks gives particularly rich system level information to understand the functional organization of living cells [1-6]. Determination of network modules (i.e. network groups, or communities) became a focal point of the analysis of network topology leading to more than a hundred independent methods to solve this challenging problem. In protein-protein interaction networks tight modules are corresponding to large protein complexes. However, more extensive, pervasively overlapping modules detected by recent methods, including ours, revealed a deeper insight to the multi-functionality of cellular proteins [7-9]. Despite of the widespread studies on network modules, the overlaps of interactome modules have not been studied yet in detail.

Network dynamics received an increasing attention in recent years. The stress response, inducing a genome-wide up- and down-regulation of gene expression after an abrupt environmental stimulus, is a particularly good model of the reorganization of cellular networks, where the observed changes have a paramount importance in survival, adaptation and evolution [10-13]. Yeast is an appropriate model organism for studying the system-level changes after stress, since we have an extensive knowledge on the organization of the yeast PPI network (interactome) [14-17], and stress-induced changes in the yeast gene expression pattern have also been studied in detail [18,19]. Despite of major interest in key biological examples of network dynamics, changes of protein-protein interaction networks in stress have not been analyzed yet.

There are two main ways to integrate gene expression data with interactome, identifying active subnetworks [20-22] or analysing the whole interactome under genomic responses [15,16,23]. In the current study we used the latter approach and assessed the changes of the yeast interactome after the archetype of stress, heat shock. Upon heat shock the yeast PPI network became a much 'larger world': heat shock induced a close to 5-fold increase in the weighted diameter and a significant, but partial disintegration of the modular structure of the yeast interactome. The decrease of inter-modular protein-protein contacts may enable a 'post heat shock' re-integration of the yeast protein-protein interaction network communities, where the slightly different inter-modular contacts may provide a cost-efficient adaptation response to the changed environment.

# RESULTS

**Global changes of the yeast interactome topology in heat shock**

To investigate the changes of the yeast interactome topology in heat shock, a well-characterized form of stress, we calculated the weight of each protein-protein interaction both in resting state and after heat shock. We used the physical protein-protein interaction subset of the BioGRID database [24], combining the benefits of this comprehensive, literature curated database with the more reliable, direct relationship of physical interactions. (We also extended our studies to a high-confidence PPI dataset, and found similar results as described in Methods.) Link weights of both basal state and heat shocked yeast cells were approximated using mRNA levels, since large-scale, complete datasets for protein abundances are currently missing (see Methods). We chose heat shock, as the form of stress we studied in detail, since it is considered to be a 'severe stress', where a good correlation between the transcriptome and the translatome has been demonstrated [25]. Interaction weights of the yeast PPI network were generated by averaging of the mRNA abundances of the two interacting proteins. Baseline and 15 min, 37°C heat shocked mRNA levels were obtained from the Holstege- [26] and Gasch-datasets [19], respectively, as described in the Methods section in detail.

The distribution of interaction weights showed a significant decrease upon heat shock (Figure 1 of Text S1; Wilcoxon paired test, $p<2.2*10^{-16}$). To interpret this change we note, that the PPI networks of 'resting' and heat shocked yeast cells had the same links. However, the two interactomes had a largely different weight structure due to the differences in mRNA expression pattern upon heat shock. Table 1 shows a few main attributes of the interactome topology of unstressed and heat shocked yeast cells. In agreement with the significant change in weight distribution, the median weight of interactions had a 14% decrease in heat shock yeast cells. Interestingly, in unstressed yeast cells larger mRNA levels were predominantly associated with larger unweighted degrees, while in heat shocked yeast cells larger mRNA levels were predominantly associated with lower unweighted degrees. Thus, heat shock induces a shift of connection weights from hub-like proteins to non-hubs (see Figure 2 of Text S1), which may indicate a partial uncoupling of the local segments of yeast interactome upon heat shock.

The most remarkable change was the close to 5-fold (491%) increase of weighted diameter (Table 1). This was a rather suprising finding, which reflected that the interactome became a much 'larger world' after heat shock. The increase of weighted diameter was accompanied by shift in the distribution of weighted shortest path lengths (based on Dijkstra's algorithm [27]) towards longer paths, causing a significant difference (Wilcoxon paired test, $p<2.2*10^{-16}$). Similarly to these findings, the average weighted shortest path length also showed a large increase (47.1 in unstressed versus 263.8 in heat shocked yeast cells). The distribution of 'effective weighted degrees' showed a scale-free like pattern, and a significant shift towards lower degrees after heat shock (Figure 3 of Text S1; Wilcoxon paired test, $p<2.2*10^{-16}$). We note, that the 'effective weighted degree' captures the total number of fractional weighted connections of a node to another (see Methods and [8] for details). The shift towards lower weighted degrees was reflected by the decrease in both the median weighted degree and the number of hubs (14% and 22% decrease, respectively; Table 1).

The decrease of median interaction weights, median weighted degree and number of hubs indicated that heat shocked yeast cells developed a generally less intensive, 'resource-sparing' interactome. The 'resource-sparing' character is in agreement with the close to 5-fold increase of weighted diameter showing that the yeast interactome preferably 'spares' the shortcuts, and becomes much less integrated upon stress.

Visual inspection of stress-induced changes of the entire yeast interactome is of limited value, since the multitude of interactions makes the comparison difficult. However, there are comprehensible subnetworks allowing an easy, pair-wise assessment. We show the subnetworks of the strongest and weakest links on Figure 1. The subnetwork of strongest links (cf. Figures 1A and 1B) of unstressed yeast cells contained a highly connected ribosomal protein complex (see Figure 1A, inset) and an additional center of carbohydrate metabolism (see Figure 1A, right bottom). Both centers are crucial for the fast cell divisions characteristic to unstressed yeast cells. Please note that the number of links is the same in both panels. Therefore, the link-density of the two major centers is much larger than the apparent density shown on the figure. Upon heat shock several locally dense regions appeared, which were centered on heat-shock proteins (see circles on Fig. 1B). This structure showed a re-organization of the interactome around proteins crucial in cell survival and recovery including dehydrogenases, proteins of glucose metabolism, a key player of protein degradation (polyubiquitin), as well as the molecular chaperones, Hsp70 and Hsp104 as detailed in the legend of Figure 1. The subnetwork of network-integrating weakest links [1-3,6,28] had a uniform link-density in basal state (Fig. 1C). After heat shock a very densely connected twin-centre of nucleolar proteins emerged (see the right side of Figure 1D) responsible for rRNA processing and ribosome biogenesis (~80 and ~90% of genes by GO term, respectively; $p<10^{-30}$ in both cases by hypergeometric test). This is in agreement with the key role of nucleolar protein complexes in cell survival [29]. In these representations the unstressed yeast interactome was closer to an organization resembling to the flat, dense, dark and low-lying stratus clouds, whereas the interactome after heat shock was closer to a multifocal structure, resembling to puffy and white cumulonimbus clouds. In former studies 'stratus' and 'cumulus' forms were described as alternative structures of the general form of yeast interactome [30]. Stratus- and cumulus-type organizations may be differing topology classes in many types of networks, such as in protein structure networks as we proposed recently [31].

In summary, the general network parameters suggested a partial disintegration of the interactome of heat shocked yeast cells represented by the large increase in weighted diameter (Table 1), and by the emergence of a cumulus-like global organization of the subnetworks of strongest and weakest links (Figure 1). Interestingly, metabolic networks of the symbiont, *Buchnera aphidicola* [32] and the free-living bacterium, *Escherichia coli* (Figure 4 of Text S1) displayed similar patterns like the interactomes of unstressed and heat shocked yeast cells. Metabolic pathways of *B. aphidicola* (Figure 4A of Text S1) showed a rather compact organization similar to a 'stratus-type' structure, whereas *E. coli* (Figure 4B of Text S1) had a more multifocal structure similar to a 'cumulus-type' network. The latter, cumulus-like structure may show that adaptation to a variable environment resulted in a multifocal pathway structure of *E. coli*, while the stratus-like structure of the *B. aphidicola* metabolism may be a consequence of a more stable environment. These assumptions are supported by the larger modularity of metabolic networks in organisms living in variable environment than that evolved under more constant conditions [33].

**Changes of the modular organization of the yeast interactome in heat shock**

After our first results suggesting a partial disintegration of the yeast interactome in heat shock exemplified by the increased network weighted diameter and the emergence of a multifocal-like structure of the subnetworks of strongest and weakest links, we examined the heat shock-induced changes of yeast PPI network modules. For the determination of yeast interactome modules we used our recently developed ModuLand framework [8], since it detects pervasive overlaps like other recent methods [34], and therefore gives a more detailed description of PPI network modules than other modularization techniques [8,34]. Moreover, the ModuLand method introduces community centrality, which is a measure of the overall influence of the whole network to one of its nodes or links. Community centrality enables an easy discrimination of module cores, containing the most central proteins of interactome modules, and makes the functional annotation of PPI network modules rather easy [8]. These modular cores are the hill-tops of the 3D representation of the interactome on Figure 2. On the figure the horizontal plane corresponds to a conventional 2D network layout of the yeast interactome, while the vertical scale shows the community centrality value of yeast proteins. Functional annotations of the most central interactome modules are listed in Tables 1 and 2 of Text S1. In the unstressed condition (Figure 2A) the central position was occupied by two ribosomal modules showing the overwhelming influence of protein synthesis on yeast cellular functions in exponentially growing yeast cells. Though this module pair was overlapping, their cores were different. Moreover, upon heat shock the two ribosomal modules showed different alterations. The third central module contained proteins of carbohydrate metabolism reflecting the importance of energy supply in yeast growth and proliferation. The additional modules recovered several modules identified before (e.g. the proteasome, ribosome biogenesis and the nuclear pore complex, see [8]). The larger functional diversity of the modules here than that obtained in our preliminary investigations using a much smaller, un-weighted dataset [8] showed the advantages of using a large dataset and interaction weights.

In contrast with the unstressed situation, the ribosomal modules had a much smaller community centrality upon heat shock (Figure 2B), which is in agreement with the inhibition of translation after heat shock. The carbohydrate metabolism module kept its central position (Tables 1 and 2 of Text S1). A novel central module emerged containing proteins involved in the regulation of autophagy, a key process in cellular survival. Several other interactome communities also increased their community centrality, such as modules of heat shock proteins containing several major molecular chaperones and their co-chaperones (e.g.: Sti1, Hsp70, Hsp82 and Hsp104), which all play a key role in sequestering and refolding misfolded proteins after heat shock. Another module of growing centrality was the trehalose synthase module providing an important chemical chaperone for yeast survival (Table 2 of Text S1). Finally, a module of negative regulators of cellular processes (such as that of Bhm1 and Bhm2) also gained centrality (Table 2 of Text S1), exemplifying the energy-saving efforts of the yeast cell in heat shock. The more multifocal modular structure of the yeast cell after heat shock (Figure 2B) compared to the more centralized, compact modular structure of resting cells (Figure 2A) is in agreement with the partial disintegration of the yeast interactome suggested by the increasing weighted diameter (Table 1) and changes of subnetworks containing the strongest and weakest links (Figure 1).

**Partial decoupling of interactome modules in heat shock**

To analyze the changes of yeast interactome modules after heat shock further, we compared the modular distribution of proteins in unstressed and heat shocked yeast cells. Figure 3A shows the cumulative distribution of the 'effective number of modules'. The 'effective number of modules' measure efficiently captures the cumulative number of all modular fractions, where a protein belongs to (see Methods and [8] for details). After heat shock yeast proteins belonged to a significantly fewer number of interactome modules (Wilcoxon paired test, $p < 2.2*10^{-16}$). In other words this means that modules of the yeast interactome had a smaller overlap after heat shock than in the unstressed state, since there were less proteins belonging to multiple modules, i.e. modular overlaps.

Assessing the modular structure one level higher, where modules were treated as elements of a coarse-grained network [8], we compared the effective degree of modules of unstressed and heat shocked yeast cells (Figure 3B). The effective degree captures the total number of fractional weighted connections of a module to another (for details, see Methods). Upon heat shock interactome modules were connected to significantly smaller number of other modules (Mann-Whitney U test, p=0.02299). Since a link between modules is related to the overlap between them ([8], for details see Methods), the decrease of inter-modular contacts upon heat shock reflects once again a smaller overlap between the interactome communities.

The decrease of modular overlap was similar in other stress conditions (e.g. in oxidative stress, reductive stress, osmotic stress, nutrient limitation, see Figure 5 of Text S1), although the heterogeneity of these conditions did not allow to create a coherent picture in every details. The partial decoupling of the interactome modules of stressed yeast cells (Figure 3) is in agreement with the increase of weighted network diameter (Table 1) and with the appearance of a larger multifocality in both the subnetworks of strongest and weakest links (Figure 1), as well as in the 3D image of modular structure (Figure 2). All these findings show a partial disintegration of the yeast interactome upon heat shock.

**Heat shock-related proteins as integrators of the partially decoupled yeast interactome**

Prompted by our data showing a partial disintegration of the yeast interactome after heat shock, we became interested to assess those proteins, which preserve the residual integration of the interactome upon heat shock. First, we assessed the community centrality changes of yeast proteins after heat shock, since high community centrality values characterize those yeast proteins, which receive a large influence from others [8], and thus integrate the responses of the yeast interactome. As a second step, we studied the bridges, i.e. the inter-modular proteins playing a key role in the remaining connection of interactome modules after heat shock.

Figure 4 shows the comparison of the community centrality values [8] of yeast proteins before and after heat shock highlighting five markedly different behaviors. Group A proteins increased their community centrality upon heat shock, Groups B and C contain proteins, which had a continuously high community centrality, while those proteins, which decreased their community centrality are in Group D. Finally, Group E proteins had a continuously low community centrality. Table 3 of Text S1 lists the proteins of the various groups of Figure 4 with their name and functional annotation. Proteins increasing their community centrality (Group A) upon heat shock included major molecular chaperones sequestering, disaggregating and refolding misfolded proteins (Hsp42 and Hsp104), as well as stabilizing cellular membranes (Hsp12) [35]. Group A proteins were also involved in stress signaling and in stress response regulation (e. g. Psr2 phosphatase, Rsp5 ubiquitin ligase) [36,37], in autophagy regulation (Tor1, Tor2), in the reorganization of the cytoskeleton (Las17 actin assembly factor) [38] and also in yeast carbohydrate metabolism (Glk1 glucokinase, Hxt6 and Hxt7 glucose transporters). These proteins were all heat shock proteins, since they showed increased mRNA expression upon heat shock. Yeast proteins with continuously high community centrality (Group B) included ubiquitin, a ribosome associated, constitutive form of Hsp70 and several key enzymes of carbohydrate metabolism. Proteins having a high, but decreasing importance upon heat shock (Group C) were constituents of the ribosome. Importantly, enzymes and proteins involved in pre-rRNA processing, thus in the synthesis of new ribosomes, showed a large decrease in their community centrality and formed a major part of Group D. These changes reflected the down-regulation of protein synthesis and cell proliferation, which are hallmarks of the heat shock response. Group E proteins with a continuously low importance included several proteins with yet unknown functions, which is understandable knowing the minor role of these proteins both in unstressed and heat shocked yeast cells.

In summary, chaperones, proteins of stress signaling and other heat shock proteins, redirecting yeast carbohydrate metabolism in heat-shock, became key players in the residual integration of yeast protein-protein interaction network after heat shock. On the contrary, those proteins, which had been major integrators of the non-stressed yeast interactome (such as proteins of the ribosome or ribosome synthesis) lost their integrating function, and contributed to the partial, modular disintegration of yeast interactome after heat shock.

Next, we selected Group A through C proteins as they had large community centrality value in heat shocked conditions, and examined their localization in the subnetwork of the yeast interactome containing the strongest links (Figure 5). Considering that Group A proteins had low community centrality values in unstressed condition, it is not surprising that only one of Group A protein was visible in the subnetwork containing the top 4% of strongest links (Figure 5A). Group A proteins (small → large community centrality) appeared as nodes having strongest links, and occupied rather dispersed locations after heat shock (Figure 5B). Group B proteins (large → large community centrality) were accumulated in one of the two alternative centers of the subnetwork in unstressed condition, and became more dispersed after heat shock (cf. Figures 5C and 5D). Group C proteins (extra large → large community centrality) occupied the other alternative center, the dense core of the subnetwork in unstressed yeast cells, while, similarly to the other groups, they became more dispersed after heat shock (cf. Figures 5E and 5F).

In summary, proteins with large community centralities had rather condensed positions in the interactomes of unstressed yeast cells, while they occupied more scattered, dispersed positions after heat shock. This reflects well the key role of the proteins with large community centralities to maintain the integration of the cumulus-type, multifocal interactome of heat shocked yeast cells at multiple positions.

As a first inquiry to assess the role of bridges in the maintenance of interactome integrity after heat shock, we highlight a group of four proteins (Table 2; Hsp42, Hsp70, Hsp104 and glycogen phosphorylase). These proteins, beyond their very remarkable increase in community centrality values, were the only proteins, which had a parallel increase in their modular overlap upon heat shock (where the latter was defined as the effective number of their modules, the measure used already in Figure 3A). We note that this behavior was peculiar, since the modular overlap had a general *decrease* after heat shock (see Figure 3). Therefore it was plausible to claim that the 4 proteins listed in Table 2 were not only central, but also behaved as bridges, connecting parts of the partially disintegrated interactome after heat shock. It is noteworthy that 3 out of the 4 proteins are molecular chaperones (Hsp42, Hsp70, Hsp104), while glycogen phosphorylase is a key enzyme of energy mobilization, a necessity in stress. This finding is in agreement with the results of previous studies and assumptions [39,40].

As a second inquiry to study the role of bridges in the interactome of unstressed and heat shocked yeast cells, we examined changes of bridgeness of yeast proteins. Figure 6 plots the bridgeness of yeast proteins before and after heat shock. Bridgeness was defined as before [8], involving the smaller of the two modular assignments of a node in two adjacent modules summed up for every module pairs. This value is high, if the node belongs more equally to two adjacent modules in many cases, i.e. it behaves as a bridge between a single pair, or between multiple pairs of modules. Such bridging positions correspond to saddles between the 'community-hills' of the 3D interactome community landscape shown on Figure 2. Note that the bridgeness measure characterizes an inter-modular position of the node between adjacent modules, while the modular overlap measure reveals the simultaneous involvement of the node in multiple modules.

The highlighted zones of Figure 6 show that the importance of 9 bridges increased, that of 7 bridges remained fairly unchanged, while the importance of only 3 bridges decreased upon heat shock. The increase of the number of key bridging proteins upon heat shock shows the increased importance of a few interactome-intergating proteins after stress (a very strong tendency for a significant change, with p=0.051 by Mann-Whitney U test, between the highlighted bridges of Figure 6 having a value larger than 10). The position of the 7 persistently high bridges and the 9 heat shock-induced bridges in the subnetwork of the yeast interactome containing the strongest links is shown on Figure 6 of Text S1. Bridges appeared in this subnetwork at a larger ratio (31% compared to 69% before and after heat shock, respectively), and were re-organized to more inter-modular positions in the interactome of the strongest links after heat shock (Figure 6 of Text S1). Name and function of key bridges are listed in Table 4 of Text S1. The 5 bridges present in both the unstressed condition and after heat shock in the strongly linked subnetwork were Srp1, Yef3, Smt3, Ubi4 and Med7, key proteins of nuclear transport, transcription, translation and protein degradation complexes, respectively. The 6 additional bridges appearing only after heat shock in the strongly linked subnetwork were Whi3, Rpn3, Rsp5, Cbk1, Hek2 and Srs2, key proteins of protein degradation, DNA repair, mRNA sequestration and metabolism, respectively: all essential processes for cell survival in stress.

In summary, a rather interesting, complex picture emerged on interactome changes of heat shocked yeast cells. On one hand, the interactome developed a decreased integrity apperaring at several hierarchical levels of the local to global topology. The most remarkable change of all these was the heat shock-induced partial uncoupling of interactome modules. On the other hand, the remaining inter-modular connections remained or became enforced by a few key proteins involved in cell survival.

## DISCUSSION

The major findings of the current paper are the following: heat shock induces i.) an increase in the weighted diameter of yeast protein-protein interaction network (Table 1); ii.) subnetworks of strongest and weakest links as well as the modular structure show a more multifocal appearance upon heat shock (Figures 1 and 2); iii.) modules became partially decoupled in heat shock (Figure 3); and finally, iv.) a few, selected, inter-modular proteins help the integration of the partially uncoupled interactome of heat shocked yeast cells (Figures 4 to 6).

A minor part of our findings was rather obvious. As an example of this: it is more-less expected that many heat shock-induced proteins will have a larger community centrality, since they have an increased weight of their interactions (Figure 1 of Text S1), and therefore, may receive a larger influence of other interactome segments. However, the partial disintegration of the yeast interactome after heat shock is by far not an obvious consequence of heat shock-induced mRNA changes, but a highly non-trivial adaptation to stress at the system level. It is important to note, that this major finding, the partial disintegration of yeast interactome after heat shock, appeared at several levels on network topology. At the very local level, a significant decrease was observed in the weighted degrees upon heat shock (Table 1; Figure 3 of Text S1). At the mesoscopic level a remarkable and highly robust decrease of modular overlaps occurred (Figure 3). At the global scale, a close to 5-fold increase of the weighted network diameter was observed (Table 1.). All these changes point to the same direction and suggest that a more 'sparing' contact structure develops upon heat shock allowing a better isolation and discrimination of cellular functions. The heat shock-mediated isolation and discrimination of cellular functions is also reflected by the change in the structure of strongest links (cf. Figures 1A and 1B), where a large number of disjunct network centres develop, and became connected by a few strong links after heat shock (Figure 1B), as opposed to a large density of strong links in a few centres in unstressed yeast cells (see Figure 1A, where the density is so large that it can not be readily visualized even in the magnified inset).

The observed findings were in a way indirect. Regretfully, no direct PPI network data exist for heat shocked cells, including yeast. Therefore, we had to calculate the yeast interactome weights after heat shock from mRNA data. As we noted earlier, this approach was justified by the finding that heat shock is a severe form of stress, where transcriptional and translational changes are better coupled [25]. Protein levels are also regulated by protein degradation. Though large-scale data on yeast protein half-lives exist [41], even these data cover only a part of the yeast genome, and their modification in heat shock is not known. Despite of these shortcomings of exact system level data in heat shock, the robustness of our major finding, the partial uncoupling of yeast interactome modules, suggests that the phenomenon we observed is a real, *in vivo* response of yeasts cells to heat shock.

The interactome modules of unstressed yeast cells defined in this paper correspond to the results of other modularization methods. When comparing our results with those obtained by the MCODE method [42] and of another method based on semantic similarity [43], the size of predicted complexes were different, but good functional matches could be identified. When we extended the comparison to methods detecting modules having a wide range of size, like the CNM method [44] or that of Mete et al. [45], besides some minor discrepances, nearly indentical modules were found having either a large size (like that of ribosomal assembly and maintenance) or a small size (like that of tRNA processing; data not shown). In a very interesting study Gavin et al. [14] defined core components and attachments of yeast protein complexes. Core components were constant parts, while attachments were more flexible, fluctuating parts of the protein complexes. Cores of several modules (see Table 2 of Text S1) were often highly similar to the core components Gavin et al. [14] (e.g. in case of the proteasome, mitochondrial translation or RNA polymerase complexes). Core proteins of the ribosome and carbohydrate metabolism were found to be in many attachment regions of Gavin et al. [14] (15 and 4 attachments as opposed to 0.2 and 0.8 cores on the average, respectively). This is in agreement with our current results showing that these proteins have an extremely high community centrality, i.e. accommodate a large influence of multiple interactome segments.

Our study provides the first detailed comparison of the interactome structure before and after heat shock. However, there were a few studies, which contained a part of this information directly, or indirectly. Valente and Cusick [16] mapped the modular structure of unstressed yeast cells, and (assuming that the structure is invariant) determined which modules are up- and downregulated under heat shock. They found several modules with similar functions to those of the unstressed cells detected in our study (e.g. ribosomes, proteasomes and complexes involved in cell cycle control, or the organization of the chromosome and cytoskeleton). The heat shock-induced changes were also similar, showing a high similarity of downregulated modules (e.g. those responsible for ribosomal function, or chromosome organization). The upregulated modules were partially consistent with our results (cell cycle control) with the exception of the proteasome and cytoskeleton organization complex. In these two exceptions we detected a central role of these two modules in the unstressed condition already, which made the detection of their further upregulation difficult. Another comparison arose from the study of Komurov and White [15], who identified static and dynamic modules. Very interestingly, modules that were found only in unstressed or heat shocked conditions in our study corresponded to their dynamic modules (regulation of intracellular pH, proteasome, ribosome biogenesis, trehalose biosynthesis). Wang and Chen [46] developed an integrated framework of gene expression profiles, genome-wide location data, protein-protein interactions and several databases to study the yeast stress response. Their study shows the system-level *mechanism* of the yeast stress response highlighting the major transcription factors of this process. The study complements ours describing stress-induced *consequences* at the systems level. The results of Wang and Chen [46] demonstrated a large degree of general similarity of various stress responses in yeast (among others showing that 136 out of 190 transcription factors are conserved in osmotic, oxidative and heat shock), which is in agreement with the similarity of interactome-level changes of network topology after various types of stresses we observed in yeast (Figure 5 of Text S1).

Our results may put the 'stratus/cumulus debate' [30,47,48] in the new contextual framework of cellular dynamics. Our findings showed that the unstressed yeast interactome resembles more to a stratus-type, whereas the heat shocked (stressed) interactome resembles more to a cumulus-type organization. This indicates that the stratus and cumulus interactome conformations may not be as antagonistic as thought before, and none of them may be a clear artifact. Our results suggest that both network conformations may occur *in vivo*, and may characterize different states of the organism. Regretfully no quantitative measures for this structural feature have been defined so far. This will be a subject of further interesting studies.

Our earlier surveys of the literature anticipated a stress-induced decrease in the number and weights of interactions, as well as the decoupling of network modules. Chaperones were hypothesized to play a major role in the coupling/decoupling processes, since they occupied a more central position during stress, and their occupation by damaged, misfolded proteins after heat shock led to a release of their former targets. This phenomenon was termed by us as 'chaperone overload' [39,49]. Our recent results support these previous considerations. Moreover, the present findings considerably extend the earlier assumptions showing the details of the heat shock-induced partial disintegration of the yeast interactome.

What may be the reasons, which make a partial disintegration of the interactome an evolutionarily profitable response for yeast cells after heat shock? i.) The decreased number and weights of interactions may be regarded as parts of the energy saving mechanisms, which are crucial for survival. The specific decrease of inter-modular contacts may 'slow down' the information transfer of stressed cells, which is a further help to save energy. ii.) The increased weighted diameter and the partially decoupled modular structure of the interactome may localize harmful damages (e.g. free radicals, dysfunctional proteins), and thus may prevent the propagation of damage. iii.) Dissociation of modules may help the mediation of 'intracellular conflicts', e.g. opposing changes in protein abundance and dynamics in stress. iv.) The appearance of a more pronounced modular structure may allow a larger autonomy of the modules. This is beneficial, since more distinct functional units may work in a more specialized, more effective way, and at the same time may also explore a larger variety of different behavior, since in their exploratory behavior they are not restricted by other modules to the extent than before stress. The larger autonomy of modules increases both the efficiency and learning potential of the cell sparing additional energy.

The observed partial disintegration of the yeast interactome after heat shock is most probably only transient. The partial de-coupling of the interactome modules is presumably followed by a re-coupling after stress, which not only restores a part of the original, denser inter-modular connections, but may also build novel inter-modular contacts, giving a structural background to the adaptation of the novel situation [39,40,50,51]. This brings a novel perspective to those proteins, which help to maintain the integration of the yeast interactome during heat shock, since some of these inter-modular proteins may play a role in the adaptive reconfiguration of PPI network as a response to the changed environment. The presence of 3 major chaperones among those 4 proteins, which increased their inter-modular overlap upon heat shock (Table 2), supports this assumption, since chaperones are well-known mediators of cellular adaptation in stress and during evolution [39,49].

The decrease of modular overlap was similar in other stress conditions (e.g. in oxidative stress, reductive stress, osmotic stress, nutrient limitation; see Figure 5 of Text S1), although the heterogeneity of these conditions inhibited to create a coherent picture in every details. Prompted by the generality of stress-induced partial disintegration of the yeast protein-protein interaction network, and by the generality of the beneficial reasons behind these changes, we were interested to see, whether similar changes may occur in other complex systems. Bagrow et al. [52] showed that network failures of a model system cause the uncoupling of overlapping modules before the loss of global connectivity. A similarly modular, sequential disruption of (presumably inter-modular) links was observed, when single molecules of the giant protein, titin were pulled introducing a physical stress [53]. Bandyopadhyay et al. [54] showed that while protein complexes tend to be stable in response to DNA damage in a genetic network, genetic interactions between these complexes are reprogrammed. Similarly to the changes shown on Figure 4 of Text S1, the group of Uri Alon found that networks of organisms in variable environment are significantly more modular than networks that evolved under more constant conditions [33,55]. These studies all revealed the stress-related dynamism of intermodular regions in other cellular contexts.

Looking at even broader analogies Tinker et al. [56] showed that food limitation causes a diversification and specialization of sea otters that greatly resembles to the changes of yeast interactome modules in stress. A similar increase of modularization (patchiness) was observed in increasingly arid environments suffering from a larger and larger drought stress [57]. A partial decoupling of social modules was also observed, when criminal networks faced increased prosecution [58]. A recent study detected a reorganization of brain network modules during the learning process [59]. As a far-fetched analogy, stress-induced psychological dissociation [60] may also be perceived as a partial decoupling of psychological modalities. The stress-induced uncoupling/recoupling cycle greatly resembles Dabrowski's psychological development theory of positive disintegration [61], as well as the Schumpeterian concept of "creative destruction" describing long-term socio-economic changes [62]. In agreement with this general picture, Brian Uzzi and co-workers [63] recently showed that brokers shift their link-structure of instant messaging from weak to strong ties under the initial phase of crisis-like events at the stock-exchange, which may reflect a partial de-coupling of weakly linked broker-network modules together with an increase of strong link-mediated intra-modular cohesion. Estrada et al. [64] proposed a model, where communicability and community structure of socio-economic networks are affected by external stress (e.g. by social agitation, or crisis). They showed that community overlaps diminished with the increase of stress. Increased modularity of the banking system may be a very efficient way to prevent the return and extension of the recent crisis in economy as pointed out recently by Haldane and May [65], and as applied by the Volcker Rule in the USA. These broad analogies are supported further by the previously proposed [31] generality of the two basic network conformations, the stratus- and cumulus-like network topology observed here before and after heat shock, respectively.

In summary, the major finding of our study was that heat shock i.) induces the increase in the weighted diameter of the yeast interactome; ii.) sets up multifocality in both subnetworks and modules of the yeast interactome, as well as iii.) contributes to the decoupling of the modules of the heat shocked yeast interactome. Parallel with these changes a few remaining inter-modular connections play an enhanced, prominent role in the residual integration of the yeast interactome. Our work may provide a model of a general, system-level adaptation mechanism to environmental changes.

**METHODS**

**Yeast protein-protein interaction (PPI) networks**

The budding yeast (*S. cerevisiae*) PPI data were from the BioGRID dataset [24] (www.thebiogrid.com, 2.0.58 release), which is a freely accessible database of physical and genetic interactions. To avoid indirect interactions only the physical interactions of the database were used. These interactions (contained in the experimental system column of the database) included physical *in vitro* interactions such as biochemical activity-derived, co-crystal structure-related, far-Western, protein-peptide, protein-RNA, or reconstituted complex interactions, as well as physical *in vivo* (like) interactions, such as affinity capture mass spectrometry, affinity capture RNA, affinity capture Western, co-fractionation, co-localization, co-purification, fluorescence resonance energy transfer and two-hybrid interactions. The giant component of the obtained PPI network was used containing 5,223 nodes and 44,314 interactions. In the absence of reliable and large-scale weighted yeast protein-protein interaction data, network link weights were generated from mRNA microarray datasets as described later. We also analyzed the high-confidence PPI dataset of Ekman et al. [23], where the giant component of the network comprised 2,444 proteins and 6,271 interactions. These results were consistent with our presented findings (Figure 7A of Text S1), although the small scale of network and the nature of interactions (which were not restricted to physical interactions as our dataset), reduced the biological relevance of this latter analysis.

**Yeast mRNA microarray data**

Yeast whole-genome mRNA expression datasets were from Holstege et al. [26] (called as the "Holstege-dataset") as a reference dataset for the baseline, non-stressed yeast gene expression profile, and from Gasch et al. [19] (called as the "Gasch-dataset") measuring relative expression profiles in various stress conditions. The Holstege-dataset contained data of 5,449, while the Gasch-dataset contained data of 6,152 yeast genes, respectively. From the Gasch-dataset we selected heat shock as the archetype of stress conditions. Besides being a widely examined form of stress, heat shock is considered as a "severe stress" by Halbeisen and Gerber [25], where a good correlation between translational and transcriptional changes have been found. We analyzed the 'hs-1' condition of the Gasch-dataset (15 minutes of 37°C heat shock), where broader time series were monitored than at 'hs-2' or other heat shock conditions (the stress condition names are the same as used by Gasch et al. [19]). We performed our analysis using longer durations of 37°C heat shock (40 and 80 minutes compared to that of the 15 minutes of the "hs-1" dataset, [19]). In line with the expectations, heat shock induced gene expression was less remarkable after 40 minutes and returned close to the baseline level after 80 minutes. Therefore we performed a detailed analysis only with the 15 minutes heat shock dataset. Importantly, our major finding, the decrease of modular overlaps after stress was robust, and persisted in all heat shock conditions tested. The decrease of modular overlap was similar in other stress conditions (e.g. in oxidative stress, reductive stress, osmotic stress, nutrient limitation, see Figure 5 of Text S1), although the specificity and heterogeneity of these conditions inhibited to create a coherent picture in every details.

Although logarithmic transformations are extensively applied in the literature, we used absolute expression values. The use of absolute expression values instead of logarithmic values was in part due to the technical difficulty that after the logarithmization step negative protein-protein interaction weights would also arose that could not be interpreted. Negative weights of the logarithmized mRNA data could be avoided applying a 1000-fold increase as a rescaling correction, which is appropriate all the more, since protein levels are roughly by this magnitude higher than the corresponding mRNA levels [66]. Using this methodology, we got similar major findings as those shown in the paper (Figure 7B of Text S1). However, due to the larger number of correction steps we did not pursue this approach in detail.

**Conversion of mRNA expression data to protein-protein interaction network weights**

Weights of interactions in the PPI network were generated from the mRNA expression data in two steps. 1.) In the first step the baseline, non-stressed protein abundances were taken as the mRNA expression levels of the Holstege-dataset [26], then the baseline protein abundance values were multiplied by the relative mRNA changes of the Gasch-dataset [19], resulting in the approximated protein abundances after heat shock.

Since the Gasch-dataset contained only relative values, and therefore could not be used as a baseline-dataset, we had to use the Holstege-dataset to calculate the baseline weights of the PPI network. To check, whether our results are sensitive for baseline selection, we performed our analysis using another gene expression dataset, where time zero data were also provided [18]. This approach resulted in a similar decrease of modular overlaps (data not shown), showing that using two different datasets for mRNA abundances do not cause unexpected variability. Due to the greater ratio of missing data (~14% in baseline data and ~11% after heat shock) we did not prefer this dataset in detailed analyses. We also tried to use protein abundances instead of mRNA abundances for the unstressed condition [67,68], but due to the large amount of missing data in these data sets (>50%) we have not pursued this approach further.

When using the mRNA changes as approximations of changes in protein levels, in agreement with Halbeisen and Gerber [25], we assumed that the mRNA expression data in heat shock correlate well with protein abundance. Missing expression data for proteins in the PPI network (436 nodes total in the baseline network, less than 9% in case of the Holstege-dataset, as well as 504 nodes total in the network after heat shock, less than 10% of the Gasch-dataset) were substituted by the median expression values (0.8 in case of the Holstege-dataset, and 0.9931 in case of the Gasch-dataset), where the median was selected instead of the mean, since the distributions also contained extreme values.

2.) In the second step link-weights of the PPI network were generated by averaging of the abundances of the two proteins linked. We also tried multiplication instead of averaging that gave very similar results and provided sufficiently robust data in case of the smaller, high-confidence PPI dataset of Ekman et al. [23] (see Figure 7A of Text S1). However, we rejected this approach in case of the BioGRID dataset, as in case of this much larger dataset it resulted in a 'rougher' community landscape with more extreme changes of community centralities than averaging, which has been generally used in calculation of our data.

The use of an unweighted baseline PPI network resulted in much less consistent data due to the large difference between the homogeneity of the unweighted baseline and the inhomogeneity of the weighted heat shocked PPI networks. The physical meaning of heat shock-induced changes in gene expression is encoded precisely by the changes of link weights at the network level. This assumption makes it understandable that an unweigthed network gave false positive results in important parts of the analysis. This has two major reasons. On one hand, community centrality values are largely affected by the density of interactions. Therefore, in an unweigthed network, proteins having a high link density in their neighborhood would result in high community centrality values independently from their expression level. On the other hand, the metrics used in the analysis (e.g. overlap as the effective number of modules) are sensitive measures of fine topological changes, therefore they were largely different in the unweighted, homogenous interactome as compared to the weighted, heterogeneous interactome.

In principle, 'relative changes' of mRNA expressions could also be used for comparison (where a, say, 4-fold increase in mRNA expression of a given gene can be split to a 2-fold decrease of its baseline abundance and a 2-fold increase of its abundance after stress corresponding to the abundances of the same protein in resting and stressed yeast cells, respectively). However the use of these 'relative changes' of mRNA expression resulted in a large variability of the baseline PPI network weights (Figure 8 of Text S1). The method using the average of protein abundance values as interaction weights, we described above, gave a reliable probabilistic model, since the more abundance the associated proteins had, the more possible they interacted, and the more weight of their PPI network link possessed. Moreover, by considering the baseline expression rates, we received a more exact description of the importance of proteins in the yeast cell in both baseline and stressed conditions.

**Analysis of the modular structure of the yeast interactome**

Yeast PPI network modules were determined using the NodeLand influence function calculation algorithm with the ProportionalHill module membership assignment method of the ModuLand module determination method family described by the authors' lab recently [8]. During the post-processing of the module assignment no merging of primary modules was applied. The ModuLand method determines extensively overlapping network modules by assigning proteins to multiple modules, which reflects well the functional diversity of proteins. The ModuLand method constructs a community landscape, where the landscape height of a protein corresponds to a community centrality value showing the influence of the whole PPI network to the given protein, thus the importance of the appropriate protein in the whole yeast interactome. In fact, community centrality is a summarized value, where in the first step of the method (currently: the NodeLand influence function calculation algorithm) all increments of the influence of other proteins to the given protein are summed up. In the second step of the calculation process (currently: the ProportionalHill modules membership assignment method) proteins with locally high community centrality (corresponding to 'hills' of the community landscape, see the 3D image of Figure 2) form the core of a module of the interactome. Individual proteins are characterized by their membership assignment strength to all interactome modules. (Usually one or a few of the modules are the ones, where the protein belongs the most, while all the other modules contain the protein only marginally). With the ModuLand framework the functional annotation of modules becomes rather easy, since it can be derived from the functions of the 'core proteins' having the largest community centrality in the module. In the current work core proteins of a given module were determined as the 5 proteins having the maximal community centrality (the number of core proteins has been extended to 8 in some exceptional cases, where indicated). Comparison of the functions of proteins with lower community centralities than that of the core proteins did not change the consensus of functional annotation of modules ([8] and Table 1 of Text S1).

**Calculation of the effective degree of nodes and modules, as well as the effective number of modules**

The effective degree of nodes and modules, as well as the effective number of modules were calculated as described earlier [8]. All effective numbers refer to a set of data, where the sum is not calculated as a discrete measure, but as a continuous measure taking into account the weighted values of the data. The effective numbers were calculated using the subsequent equation: $n_i\{V[i]\} = \exp\left(-\sum_i p_i \log p_i\right)$, where data were in set $V$, $V[i]$ was the value of data $i$, and $p_i = \dfrac{V[i]}{\sum_j V[j]}$. The dataset, $V$ contained i.) in case of the effective degree of nodes the weights of the interactions of the given node as defined earlier; ii.) in case of the effective degree of modules the weights of the links of the given module to all neighboring modules as defined here later; and iii.) in case of the effective number of modules the module membership assignment strengths of the given node to all modules of the yeast interactome. The weight of the link between modules $i$ and $j$ was the sum of the node-wise calculated overlap values $O_{ij}(n)$: $W_{(i,j)} = \sum_n O_{ij}(n)$, where $O_{ij}(n)$ was proportional to the module membership assignment strengths $H_i(n)$ and $H_j(n)$, and was normalized to the community centrality as: $O_{ij}(n) = 2\dfrac{H_i(n)H_j(n)}{c(n)}$, where $c(n)$ was the community centrality of node $n$, and the factor 2 referred to that both directions between the modules have been taken into account.

**Functional categorization of proteins and modules of yeast protein-protein interaction networks**

For the functional categorization of yeast PPI network modules (see Tables 1 and 2 of Text S1), the Gene Ontology (GO) term, biological process [69] (http://www.yeastgenome.org/cgi-bin/GO/goTermFinder.pl) of the core modular proteins (as defined above) were compared. A modular GO term was assigned, if the core proteins shared a significant ($p<0.01$) amount of their GO terms. GO terms of only the most central modules were identified, since they were supposed to have a relevant role in cellular functions. The threshold was applied by the community centrality values of the most central proteins of modules (where community centrality values were greater, than 500), and this resulted in 15 or 14 modules for the unstressed or heat shocked conditions, respectively. In those exceptional cases, when the 5 core modular proteins did not result in a meaningful functional assignment (in case of 5 modules representing 17% of the 29 modules total), we extended the core-set to 8 proteins. Only 2 modules (representing 7% of the 29 modules total) were found, where none of these definitions resulted in any common assignment.

**Statistical methods**

For the statistical evaluation of data the non parametric statistical tests of the Mann-Whitney U test and the Wilcoxon paired test were applied using the R-statistical program (https://www.r-project.org) as described in the actual experiments. The hypergeometric test was performed as provided by the Gene Ontology Term Finder: http://www.yeastgenome.org/cgi-bin/GO/goTermFinder.pl.

SUPPORTING INFORMATION

*Text S1:* This supporting information contains 8 Figures, 4 Tables, as well as 8 References.

ACKNOWLEDGMENTS

The authors thank Balázs Szappanos and Balázs Papp (Evolutionary Systems Biology Group, Biological Research Center, Szeged) for the construction of the metabolic network data, and members of the LINK-Group (www.linkgroup.hu), especially Robin Palotai and Marcell Stippinger for their discussions and help.

REFERENCES

1. Barabasi AL, Oltvai ZN (2004) Network biology: understanding the cell's functional organization. Nat Rev Genet 5: 101−113.
2. Albert R (2005) Scale-free networks in cell biology. J Cell Sci 118: 4947–4957.
3. Csermely P (2009) Weak links: The universal key to the stability of networks and complex systems. Heidelberg: Springer. 410 p.
4. Gursoy A, Keskin O, Nussinov R (2008) Topological properties of protein interaction networks from a structural perspective. Biochem Soc Trans 36: 1398–1403.
5. Tsai CJ, Ma B, Nussinov R (2009) Protein-protein interaction networks: how can a hub protein bind so many different partners? Trends Biochem Sci 34: 594–600.
6. Zhu X, Gerstein M, Snyder M (2007) Getting connected: analysis and principles of biological networks. Genes Dev 21: 1010−1024.
7. Fortunato S (2010) Community detection in graphs. Phys Rep 486: 75−174.
8. Kovacs IA, Palotai R, Szalay MS, Csermely P (2010) Community landscapes: an integrative approach to determine overlapping network module hierarchy, identify key nodes and predict network dynamics. PLoS ONE 5: e12528.
9. Li X, Wu M, Kwoh CK, Ng SK (2010) Computational approaches for detecting protein complexes from protein interaction networks: a survey. BMC Genomics 11: S3.
10. Hightower LE (1991) Heat shock, stress proteins, chaperones, and proteotoxicity. Cell 66: 191–197.
11. Hong CI, Zámborszky J, Csikász-Nagy A (2009) Minimum criteria for DNA damage-induced phase advances in circadian rhythms. PLoS Comput Biol 5: e1000384.
12. Horváth I, Vígh L (2010) Cell biology: Stability in times of stress. Nature 463: 436–438.
13. Przytycka TM, Singh M, Slonim DK (2010) Toward the dynamic interactome: it's about time. Brief Bioinform 11: 15–29.
14. Gavin AC, Aloy P, Grandi P, Krause R, Boesche M, et al. (2006) Proteome survey reveals modularity of the yeast cell machinery. Nature 440: 631−636.
15. Komurov K, White M (2007) Revealing static and dynamic modular architecture of the eukaryotic protein interaction network. Mol Syst Biol 3: 110.
16. Valente AX, Cusick ME (2006) Yeast protein interactome topology provides framework for coordinated-functionality. Nucleic Acids Res 34: 2812−2819.
17. Cohen-Gihon I, Nussinov R, Sharan R (2007) Comprehensive analysis of co-occurring domain sets in yeast proteins. BMC Genomics 8: 161.
18. Causton HC, Ren B, Koh SS, Harbison CT, Kanin E, et al. (2001) Remodeling of yeast genome expression in response to environmental changes. Mol Biol Cell 12: 323−337.
19. Gasch AP, Spellman PT, Kao CM, Carmel-Harel O, Eisen MB, et al. (2000) Genomic expression programs in the response of yeast cells to environmental changes. Mol Biol Cell 11: 4241−4257.
20. Cabusora L, Sutton E, Fulmer A, Forst CV (2005) Differential network expression during drug and stress response. Bioinformatics 21: 2898−2905.
21. Guo Z, Wang L, Li Y, Gong X, Yao C, et al. (2007) Edge-based scoring and searching method for identifying condition-responsive protein-protein interaction sub-network. Bioinformatics 23: 2121−2128.
22. Ulitsky I, Shamir R (2007) Identification of functional modules using network topology and high-throughput data. BMC Syst Biol 1: 8.

23. Ekman D, Light S, Bjorklund AK, Elofsson A (2006) What properties characterize the hub proteins of the protein-protein interaction network of *Saccharomyces cerevisiae*? Genome Biol 7: R45.
24. Stark C, Breitkreutz BJ, Chatr-Aryamontri A, Boucher L, Oughtred R, et al. (2011) The BioGRID interaction database: 2011 update. Nucleic Acids Res 39: D698−704.
25. Halbeisen RE, Gerber AP (2009) Stress-dependent coordination of transcriptome and translatome in yeast. PLoS Biol 7: e105.
26. Holstege FC, Jennings EG, Wyrick JJ, Lee TI, Hengartner CJ, et al. (1998) Dissecting the regulatory circuitry of a eukaryotic genome. Cell 95: 717−728.
27. Dijkstra EW (1959) A note on two problems in connexion with graphs. Numer Math 1: 269−271.
28. Uetz P, Giot L, Cagney G, Mansfield TA, Judson RS, et al. (2000) A comprehensive analysis of protein-protein interactions in *Saccharomyces cerevisiae*. Nature 403: 623−627.
29. Bański P, Kodiha M, Stochaj U (2010) Chaperones and multitasking proteins in the nucleolus: networking together for survival? Trends Biochem Sci 35: 361−367.
30. Batada NN, Reguly T, Breitkreutz A, Boucher L, Breitkreutz BJ, et al. (2006) Stratus not altocumulus: a new view of the yeast protein interaction network. PLoS Biol 4: e317.
31. Csermely P, Sandhu KS, Hazai E, Hoksza Z, Kiss HJM, et al. (2011) Disordered proteins and network disorder in network representations of protein structure, dynamics and function. Hypotheses and a comprehensive review. Curr Prot Pept Sci. in press, available here: http://arxiv.org/abs/1101.5865.
32. Pal C, Papp B, Lercher MJ, Csermely P, Oliver SG, et al. (2006) Chance and necessity in the evolution of minimal metabolic networks. Nature 440: 667−670.
33. Parter M, Kashtan N, Alon U (2007) Environmental variability and modularity of bacterial metabolic networks. BMC Evol Biol 7: 169.
34. Ahn YY, Bagrow JP, Lehmann S (2010) Link communities reveal multiscale complexity in networks. Nature 466: 761−764.
35. Mager WH, Ferreira PM (1993) Stress response of yeast. Biochem J 290: 1−13.
36. Haitani Y, Takagi H (2008) Rsp5 is required for the nuclear export of mRNA of HSF1 and MSN2/4 under stress conditions in *Saccharomyces cerevisiae*. Genes Cells 13: 105−116.
37. Kaida D, Yashiroda H, Toh-e A, Kikuchi Y (2002) Yeast Whi2 and Psr1-phosphatase form a complex and regulate STRE-mediated gene expression. Genes Cells 7: 543−552.
38. Karpova TS, McNally JG, Moltz SL, Cooper JA (1998) Assembly and function of the actin cytoskeleton of yeast: relationships between cables and patches. J Cell Biol 142: 1501−1517.
39. Palotai R, Szalay MS, Csermely P (2008) Chaperones as integrators of cellular networks: changes of cellular integrity in stress and diseases. IUBMB Life 60: 10−18.
40. Szalay MS, Kovacs IA, Korcsmaros T, Bode C, Csermely P (2007) Stress-induced rearrangements of cellular networks: consequences for protection and drug design. FEBS Lett 581: 3675−3680.
41. Belle A, Tanay A, Bitincka L, Shamir R, O'Shea EK (2006) Quantification of protein half-lives in the budding yeast proteome. Proc Natl Acad Sci U S A 103: 13004−13009.
42. Bader GD, Hogue CW (2003) An automated method for finding molecular complexes in large protein interaction networks. BMC Bioinformatics 4: 2.
43. Cho YR, Hwang W, Ramanathan M, Zhang A (2007) Semantic integration to identify overlapping functional modules in protein interaction networks. BMC Bioinformatics 8: 265.
44. Clauset A, Newman ME, Moore C (2004) Finding community structure in very large networks. Phys Rev E 70: 066111.
45. Mete M, Tang F, Xu X, Yuruk N (2008) A structural approach for finding functional modules from large biological networks. BMC Bioinformatics 9: S19.
46. Wang YC, Chen BS (2010) Integrated cellular network of transcription regulations and protein-protein interactions. BMC Syst Biol 4: 20.
47. Batada NN, Reguly T, Breitkreutz A, Boucher L, Breitkreutz BJ, et al. (2007) Still stratus not altocumulus: further evidence against the date/party hub distinction. PLoS Biol 5: e154.
48. Bertin N, Simonis N, Dupuy D, Cusick ME, Han JD, et al. (2007) Confirmation of organized modularity in the yeast interactome. PLoS Biol 5: e153.
49. Csermely P (2001) Chaperone-overload is a possible contributor to 'civilization diseases'. Trends Genet 17: 701−704.
50. Csermely P (2008) Creative elements: network-based predictions of active centres in proteins and cellular and social networks. Trends Biochem Sci 33: 569−576.
51. Korcsmaros T, Kovacs IA, Szalay MS, Csermely P (2007) Molecular chaperones: the modular evolution of cellular networks. J Biosci 32: 441−446.
52. Bagrow JP, Lehmann S, Ahn YY (2011) Robustness and modular structure in networks. http://arxiv.org/abs/1102.5085v1

53. Kellermayer MS, Smith SB, Granzier HL, Bustamante C (1997) Folding-unfolding transitions in single titin molecules characterized with laser tweezers. Science 276: 1112–1116.
54. Bandyopadhyay S, Mehta M, Kuo D, Sung MK, Chuang R, et al. (2010) Rewiring of genetic networks in response to DNA damage. Science 330: 1385–1389.
55. Kashtan N, Noor E, Alon U (2007) Varying environments can speed up evolution. Proc Natl Acad Sci U S A 104: 13711–13716.
56. Tinker MT, Bentall G, Estes JA (2008) Food limitation leads to behavioral diversification and dietary specialization in sea otters. Proc Natl Acad Sci U S A 105: 560–565.
57. Rietkerk M, Dekker SC, de Ruiter PC, van de Koppel J (2004) Self-organized patchiness and catastrophic shifts in ecosystems. Science 305: 1926–1929.
58. Kenney M (2009) Turning to the 'dark side'. Coordination, exchange, and learning in criminal networks. In: Kahler M, editor. Networked politics: Agency, power, and governance. Ithaca: Cornell University Press. pp. 79–102.
59. Bassett DS, Wymbs NF, Porter MA, Mucha PJ, Carlson JM, et al. (2011) Dynamic reconfiguration of human brain networks during learning. Proc Natl Acad Sci U S A 108: 7641–7646.
60. Bob P (2008) Brain and dissociated mind. New York: Nova Biomedical Books. 151 p.
61. Mendaglio S (2008) Dabrowski's theory of positive disintegration. Scottsdale: Great Potential Press. 326 p.
62. Schumpeter JA (1942) Capitalism, socialism, democracy. New York: Harper. 425 p.
63. Saavedra S, Hagerty K, Uzzi B (2011) Synchronicity, instant messaging, and performance among financial traders. Proc Natl Acad Sci U S A 108: 5296–5301.
64. Estrada E, Hatano N (2010) Communicability and communities in complex socio-economic networks. In: Takayasu M, Watanabe T, Takayasu H, editors. Econophysics approaches to large-scale business data and financial crisis. Tokyo: Springer. pp. 271-288.
65. Haldane AG, May RM (2011) Systemic risk in banking ecosystems. Nature 469: 351–355.
66. Futcher B, Latter GI, Monardo P, McLaughlin CS, Garrels JI (1999) A sampling of the yeast proteome. Mol Cell Biol 19: 7357–7368.
67. Ghaemmaghami S, Huh WK, Bower K, Howson RW, Belle A, et al. (2003) Global analysis of protein expression in yeast. Nature 425: 737–741.
68. Newman JR, Ghaemmaghami S, Ihmels J, Breslow DK, Noble M, et al. (2006) Single-cell proteomic analysis of *S. cerevisiae* reveals the architecture of biological noise. Nature 441: 840–846.
69. Berardini TZ (2010) The Gene Ontology in 2010: extensions and refinements. Nucleic Acids Res 38: D331–D335.
70. Shannon P, Markiel A, Ozier O, Baliga NS, Wang JT, et al. (2003) Cytoscape: a software environment for integrated models of biomolecular interaction networks. Genome Res 13: 2498–2504.

**Table 1. Comparison of the main attributes of protein-protein interaction networks (interactomes) of unstressed and heat shocked yeast cells**

|  | Median weight[a] | Weighted diameter[b] | Median degree[a,c] | Number of hubs[d] |
|---|---|---|---|---|
| **Interactome of unstressed yeast cells** | 1.70 | 89.2 | 5.78 | 54 |
| **Interactome of heat shocked yeast cells** | 1.47 | 437.6 | 4.99 | 42 |

[a]We used median values, since distributions were not considered normal distributions. The average values of distributions showed similar results (data not shown).
[b]Weighted diameters were calculated by the igraph library as a Python extension module (version 0.5.4, http://igraph.sourceforge.net/) using Dijkstra's algoritm [27].
[c]Degree denotes the effective degree of a yeast interactome node, which was calculated as the effective number of weighted interactions of the respective node (see Methods for more details).
[d]A hub was defined as a node having more than 92 effective weighted degree (this was the effective weighted degree threshold of the top 1% of nodes having a maximal effective weighted degree in the interactome of non-stressed yeast cells).

**Table 2. Proteins having an exceptionally increasing modular overlap and increasing community centrality after heat shock**

| ORF name | Gene name | Overlap ratio[a] | Community centrality ratio[b] | Functional annotation |
|---|---|---|---|---|
| **YDR171W** | HSP42 | 1.1 | 18900 | Small heat shock protein (sHSP) with chaperone activity |
| **YPR160W** | GPH1 | 1.3 | 7500 | Non-essential glycogen phosphorylase required for the mobilization of glycogen; activity is regulated by cyclic AMP-mediated phosphorylation; expression is regulated by stress-response elements and by the HOG MAP kinase pathway |
| **YLL026W** | HSP104 | 1.1 | 27700 | Heat shock protein that cooperates with Ydj1p (Hsp40) and Ssa1p (Hsp70) to refold and reactivate previously denatured, aggregated proteins |
| **YER103W** | SSA4 | 1.4 | 6800 | Heat shock protein Hsp70 that is highly induced upon stress |

[a]Overlap denotes the effective number of yeast interactome modules that a protein is assigned to (see Methods). Overlap ratio was calculated by dividing the overlap value of the given protein in the heat shock dataset with that in the unstressed state.
[b]Community centrality values of proteins were calculated by the NodeLand influence function method [8]. Community centrality ratio was calculated by dividing the community centrality value of the given protein in the heat shock dataset with that in the unstressed state.

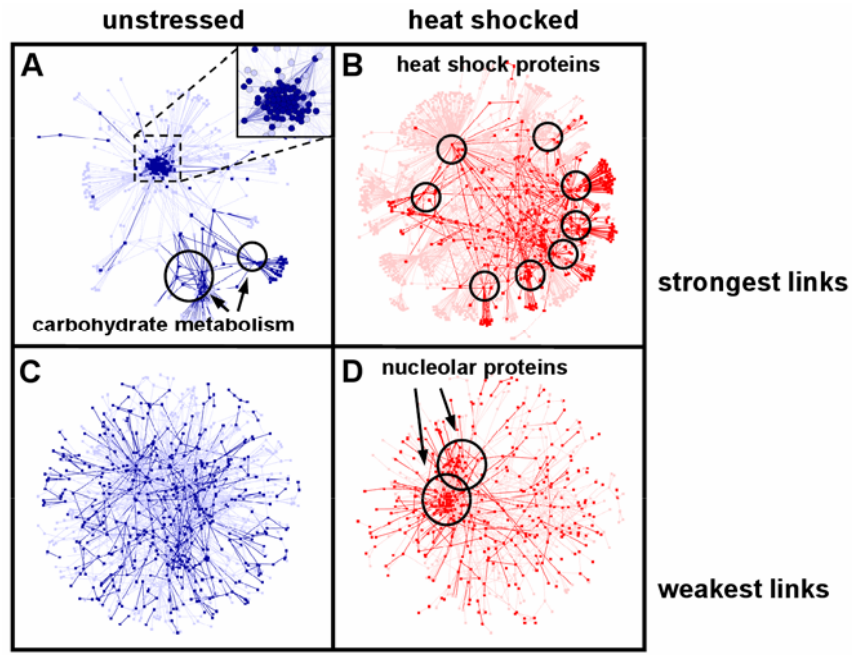

**Figure 1. Changes of yeast interactome subnetworks after heat shock.** Unstressed (panels A and C; blue) and stressed (15 min heat shock at 37°C, panels B and D; red) BioGRID yeast protein-protein interaction networks were created as described in Methods. Their subnetworks were derived from links having their interaction weights in the top (strongest links), or bottom (weakest links) 4% of all interactions. Interaction weights of the top or bottom 1% of all interaction weights and nodes having at least one of these 'top 1%' interactions were labeled with darker colors. The giant components of these subnetworks were visualized using the spring-embedded layout of Cytoscape [70]. Panels A and B. Strongest interactions of unstressed (A) and heat shocked (B) yeast interactome. The inset of Panel A shows the structure of the highly-connected ribosomal protein complex in more detail. Circles of Panel B highlight the following heat shock proteins in clockwise order starting from middle left: Hxt7, Ubi4, Tsl1, Ssa2, Hsp104, Adh1, Tdh3 and Hxk1. Panels C and D. Weakest interactions of unstressed (C) and heat shocked (D) yeast interactome.

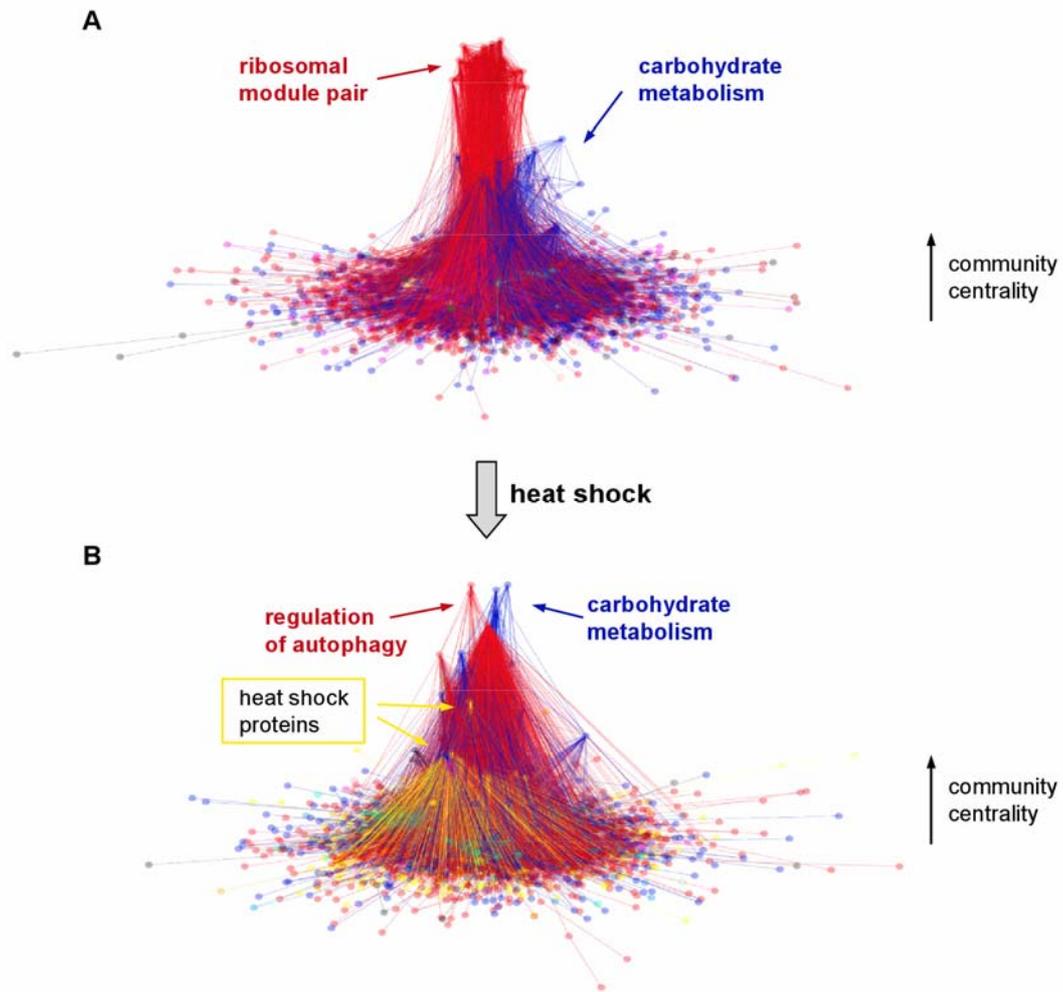

**Figure 2. Changes of the modular structure of the yeast protein-protein interaction network after heat shock.** Unstressed (panel A) and heat shocked (15 min heat shock at 37°C, panel B) yeast BioGRID protein-protein interaction networks were created as described in the Methods section. The 2D representation of yeast interactomes was visualized using the Fruchterman-Reingold algorithm. The vertical positions reflect the community centrality values of the nodes calculated by the NodeLand influence function method [8], and were plotted using a fourth root scale. Modular assignment of yeast proteins was performed by the ProportionalHill module membership assignment method [8]. Nodes were colored according to the module they maximally belong to. The functions of modules were assigned by the functions of the core modular proteins as described in the Methods section. The functional labels and the arrows had the same colors as their respective modules. Panel A. Modular structure of the unstressed yeast interactome. Two overlapping major modules had a large centrality: a ribosomal module-pair and a module representing carbohydrate metabolism. Panel B. Modular structure of the interactome of heat shocked yeast cells. The centrality of ribosomal modules decreased, which is in agreement with the diminished translation in heat shock. Besides modules of carbohydrate metabolism, upon heat shock several, formerly minor, heat shock-induced modules gained centrality, and became visible on the 3D plot. Modules related to autophagy, a key factor of the stress-response, also increased their centrality.

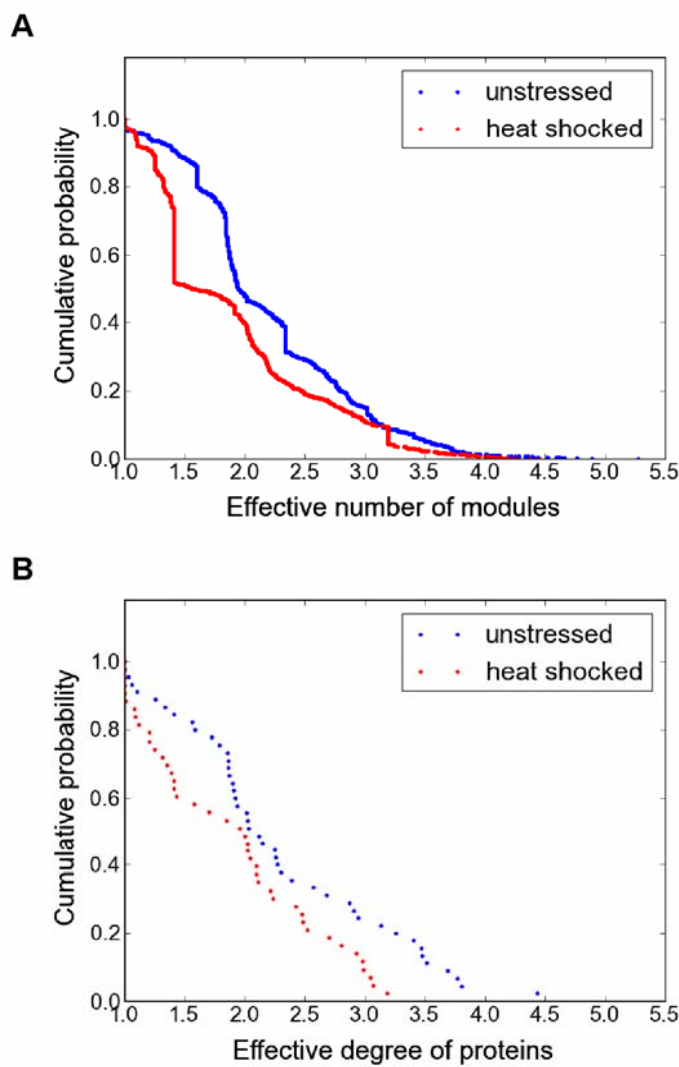

**Figure 3. Heat shock induces a partial decoupling of the modules of the yeast interactome.** Unstressed and heat shocked (15 min heat shock at 37°C) yeast BioGRID protein-protein interaction networks were created as described in the Methods section. Overlapping modules were calculated by the NodeLand influence function method combined with the ProportionalHill module membership assignment method [8] as described in Methods. Panel A. Overlap of yeast interactome modules in unstressed condition (blue dots) and upon heat shock (red dots). The overlap of yeast interaction modules was represented by the cumulative distribution of the effective number of modules of yeast proteins (for the detailed explanation of the meaning of 'effective number' describing a weighted sum of modules, see Methods). Upon heat shock the number of modules, that a yeast protein simultaneously belongs to, was significantly decreased (significance for the distribution by the Wilcoxon paired test, $p < 2.2*10^{-16}$). In other words this means that there were smaller overlaps between the interactome modules. Panel B. Cumulative distribution of the degree of yeast interactome modules in unstressed condition (blue dots) and upon heat shock (red dots). The effective degree of modules was calculated as described in Methods. Upon heat shock the cumulative distribution of effective degree of modules was significantly decreased (Mann-Whitney U test, $p = 0.02299$), which means that the protein-protein interaction network modules were less connected in heat shock than in the unstressed state.

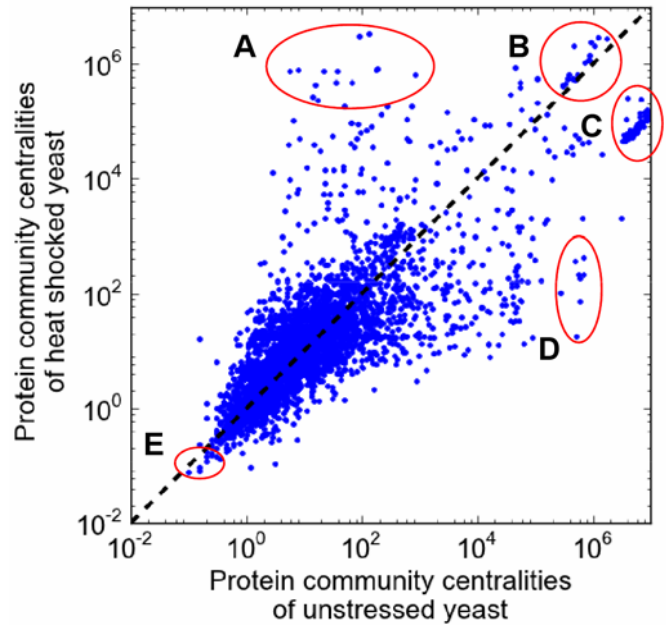

**Figure 4. Yeast proteins with altered community centrality upon heat shock.** Unstressed and heat shocked (15 min heat shock at 37°C) yeast BioGRID protein-protein interaction networks were created as described in the Methods section. Community centrality values of proteins were calculated by the NodeLand influence function method [8]. Each blue dot represents a yeast protein having its community centrality value in unstressed state plotted on the x axis, while the same value after heat shock plotted on the y axis. The 1:1 correlation is represented by the black dashed line. Five groups of proteins with extreme behavior were labeled by red circles, and indicated by letters A through E: small → large community centrality (A), large community centrality in both conditions (B), extra large → slightly smaller community centrality (C), large → small community centrality (D), small community centrality in both conditions (E). Names and functions of proteins belonging to groups A through E are listed in Table 3 of Text S1.

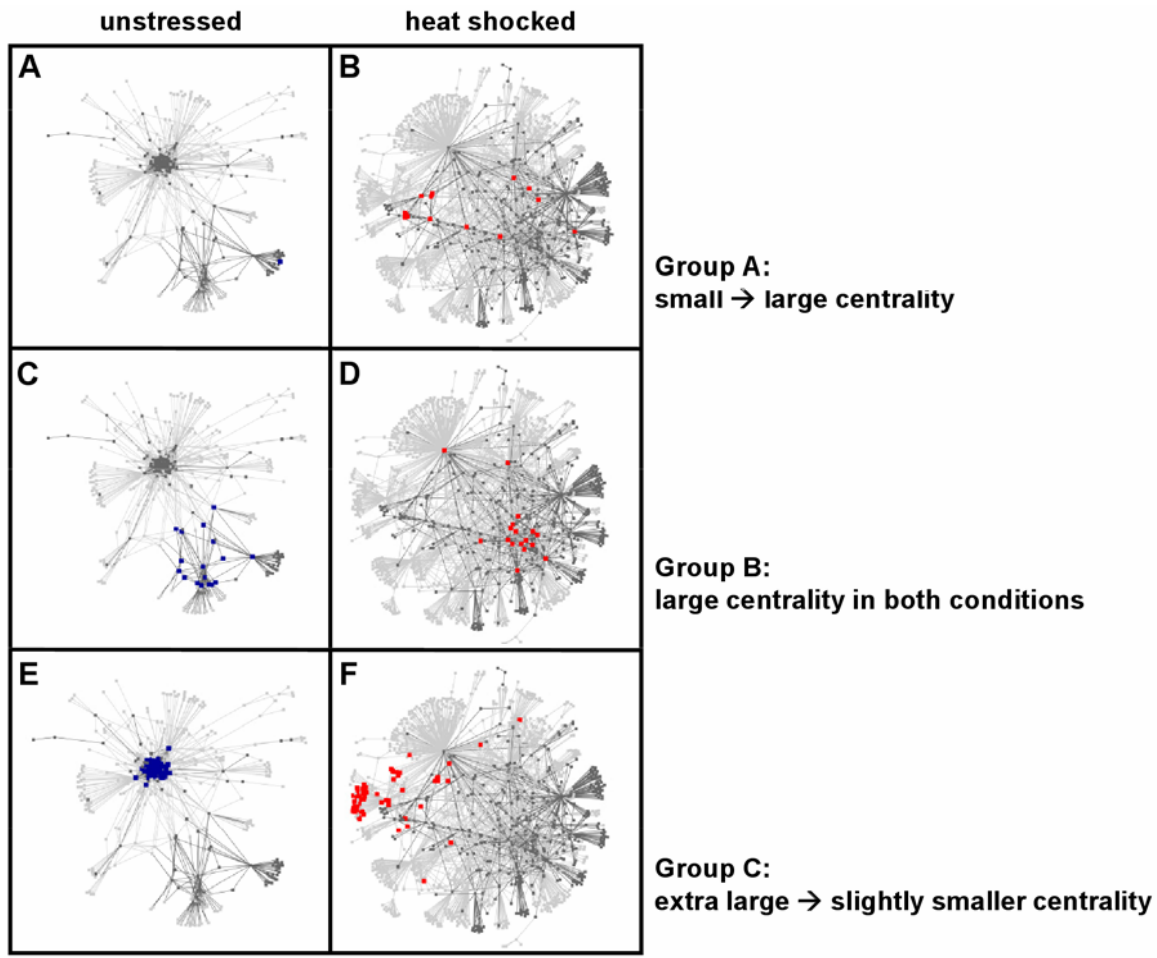

**Figure 5. Topological positions of proteins with heat shock-altered community centrality in the network of the strongest interactions of the yeast interactome.** Protein-protein interaction networks of unstressed (panels A, C and E) and heat shocked (15 min heat shock at 37°C; panels B, D and F) yeast cells were created as described in Methods. The subnetworks of their strongest links were determined and visualized as described in the legend of Figure 1. Similarly to the color-codes of Panels A and B of Figure 1, light grey colors denote the top 4%, while dark-grey colors the top 1% of interactions, respectively. Special groups of proteins with altered community centrality (Groups A through C, as described in the legend of Figure 4 and in Table 3 of Text S1) are marked with larger blue filled circles in the unstressed conditions (panels A, C and E) and with larger red filled circles in the heat shocked conditions (panels B, D and F), respectively. Panels A and B. Topological positions of 'Group A' proteins having a small → large community centrality transition upon heat shock. Only a single 'Group A' protein was among the top 4% of link weights in non-stressed condition (Panel A). 'Group A' proteins became visible and dispersed upon heat shock (Panel B). Panels C and D. Topological positions of 'Group B' proteins having large community centrality in both conditions. Proteins were condensed in one of the alternative centers before heat shock (Panel C) and became more dispersed after heat shock (Panel D). Panels E and F. Topological positions of group C proteins having an extra large → slightly smaller community centrality transition upon heat shock. Proteins were occupying the other alternative center of the subnetwork in unstressed condition (Panel E), and became dispersed upon heat shock (Panel F).

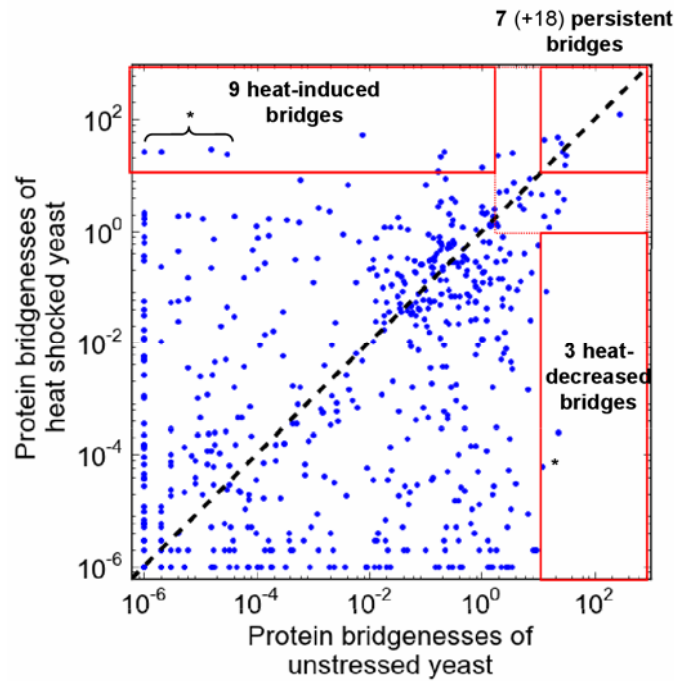

**Figure 6. Heat shock-induced changes of bridgeness of yeast proteins.** Protein-protein interaction networks of unstressed and heat shocked (15 min heat shock at 37°C) yeast cells were created and the bridgeness of their proteins was determined as described earlier [8]. Proteins having zero bridgeness values in one of the conditions were excluded from subsequent analysis. Red boxes denote those proteins, which had a large bridgeness only after heat shock (top red box containing 9 heat-induced bridges); only before heat shock (left red box containing 3 heat-decreased bridges); or were persistent bridges in both conditions (red box in top right corner containing 7 persistent bridges, as well as red dotted box in top right corner containing an additional 18 persistent, albeit less dominant bridges). Proteins were marked by asterisk, if their bridgeness induction or reduction were more than $10^5$-fold. Names and functional annotations of the bridges in the red boxes are listed in Table 4 of Text S1. The position of the 7 persistent and 9 heat shock-induced bridges in the yeast interactome containing the strongest links is shown on Figure 6 of Text S1.

# Heat shock partially dissociates the overlapping modules of the yeast protein-protein interaction network: a systems level model of adaptation

Ágoston Mihalik and Peter Csermely*

*Department of Medical Chemistry, Semmelweis University, Tűzoltó str. 37-47, H-1094 Budapest, Hungary*

*E-mail: csermely@eok.sote.hu

## Summary

This supporting information (Text S1) contains a detailed information on the distribution and variabilty of interaction weights, on correlation of mRNA abundances with unweighted degrees and on degree distributions of heat shocked yeast interactomes; a comparison of the metabolic networks of *Buchnera aphidicola* and *Escherichia coli*; additional data on the decrease of modular overlap in stresses other than heat shock and using other model parameters; as well as on the topological position of major bridges in the interactome in 8 supporting figures. The supporting information also contains the functional annotation of modules as well as the identity of major proteins with high community centrality and bridgeness values in 4 supporting tables.

# Table of contents

# Supporting Figures

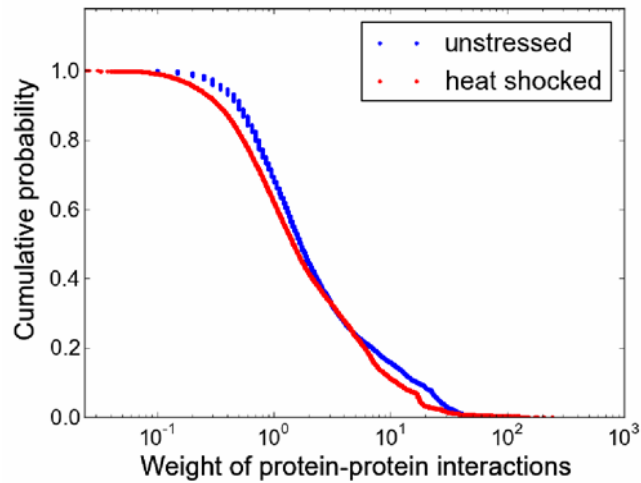

**Figure S1. Cumulative distribution of protein-protein interaction weights of unstressed and heat shocked yeast cells.** Interaction weights of yeast protein-protein interaction network (derived from the BioGRID database [1] as described in Methods of the main text) were generated by averaging of the mRNA abundances of the two interacting proteins. Unstressed and 15 min, 37°C heat shocked mRNA levels were obtained from the Holstege- [2] and Gasch-datasets [3], respectively as described in Methods of the main text. The cumulative distribution of unstressed and heat shocked interaction weights is shown using blue and red symbols, respectively. The distribution of interaction weights showed a significant shift towards lower weights upon heat shock (Wilcoxon paired test, $p<2.2*10^{-16}$).

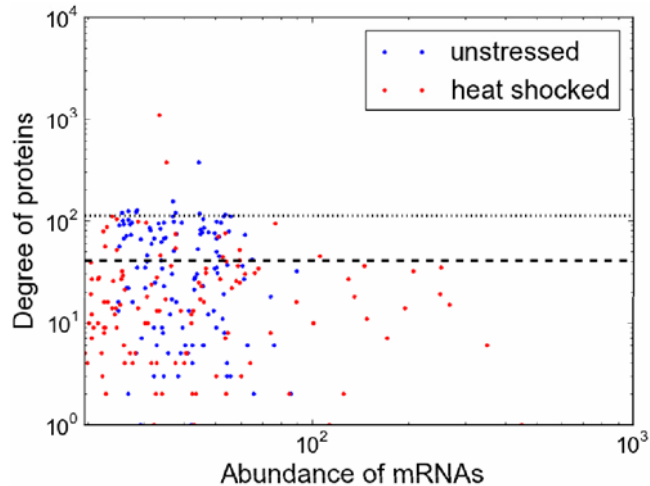

**Figure S2. Number of unweighted hubs having the top 120 mRNA levels in unstressed and heat shocked yeast cells.** Unstressed and heat shocked (15 min heat shock at 37°C) yeast mRNA levels were calculated and unweighted BioGRID protein-protein interaction networks were created as described in the legend of Figure S1 of Text S1 and in the Methods section of the main text. The top 120 mRNA abundances were selected and the unweighted degrees of the corresponding proteins were plotted as a function of their mRNA expression levels of unstressed (blue dots) and heat shocked (red dots) yeast cells. The dotted line shows the threshold of hubs set to the degree of 112 representing the top 1% of nodes having maximal unweigthed degrees in the whole interactome. The figure shows that 9 or 2 hubs were found among the proteins having the top 120 mRNA levels in unstressed or heat shocked yeast cells, respectively. The dashed line shows the top 10% of nodes, having 56 or 20 neighbor-rich nodes in unstressed or heat shocked yeast cells, respectively. Both data show that hubs associate with highly expressed mRNA (and protein) levels to a greater extent in unstressed than in heat shocked yeast cells.

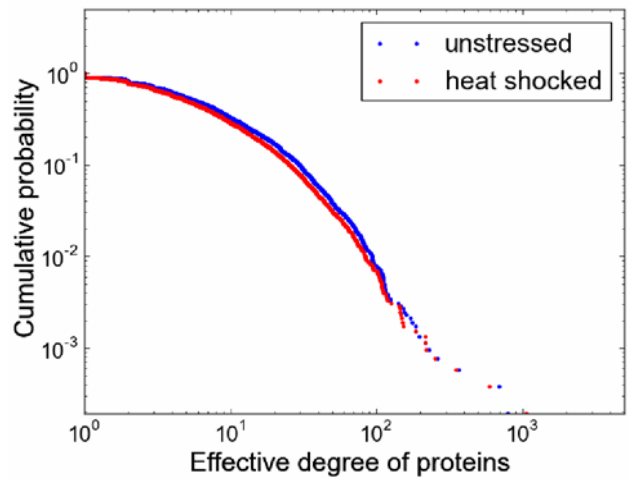

**Figure S3. Cumulative log-log distribution of weighted interactome degrees of unstressed and heat shocked yeast cells.** Unstressed and heat shocked (15 min heat shock at 37°C) yeast BioGRID protein-protein interaction networks were created as described in the legend of Figure S1 of Text S1 and in the Methods section of the main text. The effective degree was calculated as the effective number of weighted interactions of the respective node (see Methods of the main text for more details). The cumulative log-log distribution of unstressed and heat shocked effective degrees is shown using blue and red symbols, respectively. The distribution of effective degrees showed a scale-free like pattern and a significant shift towards lower degrees upon heat shock (Wilcoxon paired test, $p<2.2*10^{-16}$).

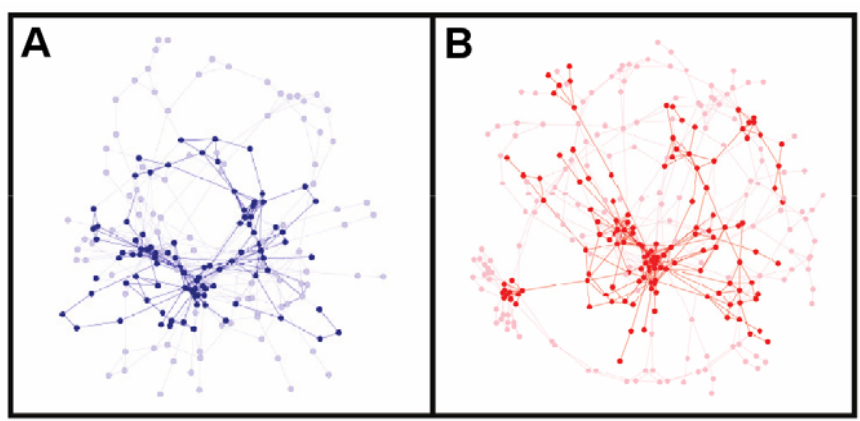

**Figure S4. Metabolic networks of the symbiont, *Buchnera aphidicola* and the free living bacterium, *Escherichia coli*.** Metabolic networks of *Buchnera aphidicola* (panel A) and *Escherichia coli* (panel B) were constructed based on the primary data of Thomas et al. [4] and Feist et al. [5], respectively. Frequent cofactors were deleted from the networks, except of those metabolic reactions, where cofactors were considered as main components. For the better comparison of networks, metabolic reactions were taken irreversible and flux balance analyses (FBA) were performed resulting in weighted networks. All flux quantities were minimized, whereas reactions non-affecting the biomass production were considered having zero flux. Weights were generated as the mean of the appropriate flux quantities in absolute value, except of the case when one of the fluxes was zero that resulted in a zero weight automatically. Subnetworks were created based on metabolic reactions having non-zero flux quantities, then giant components of the respective networks were visualized using the spring-embedded layout of Cytoscape [6]. Core reactions with weights being in the top 40% and nodes having at least one core interaction were labeled with darker colors. Panel A. Subnetwork of the metabolic network of *Buchnera aphidicola.* Panel B. Subnetwork of the metabolic network of *Escherichia coli*.

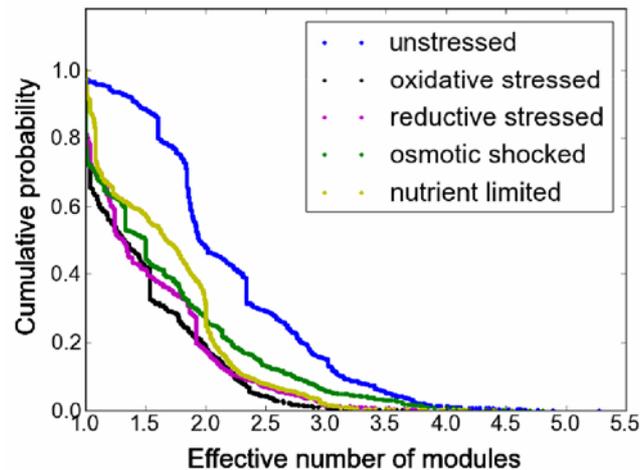

**Figure S5. Decrease of modular overlap of the yeast interactome in different stress conditions.** Protein-protein interaction weights of unstressed (blue dots), oxidative stressed (20 min menadione, black dots), reductive stressed (15 min dithiothreitol, magenta dots), osmotic shocked (15 min hypo-osmotic shock, green dots) and nutrient limited (0.5 h amino acid starvation, yellow dots) yeast BioGRID protein-protein interaction networks were created as described in the legend of Figure S1 of Text S1 and in the Methods section of the main text. Overlapping modules were calculated by the NodeLand influence function method combined with the ProportionalHill module membership assignment method [7] as described in Methods of the main text. The overlap of yeast interaction modules was represented by the cumulative distribution of the effective number of modules of yeast proteins (see Methods of the main text for more details). Upon different stress conditions (oxidative stress, reductive stress, osmotic shock and nutrient limitation) the effective number of modules, that a yeast protein simultaneously belongs to, was equally significantly decreased (Wilcoxon paired test, $p < 2.2*10^{-16}$ in all cases) similarly to the significant decrease of the same measure upon heat shock as showed in Figure 3 of the main text.

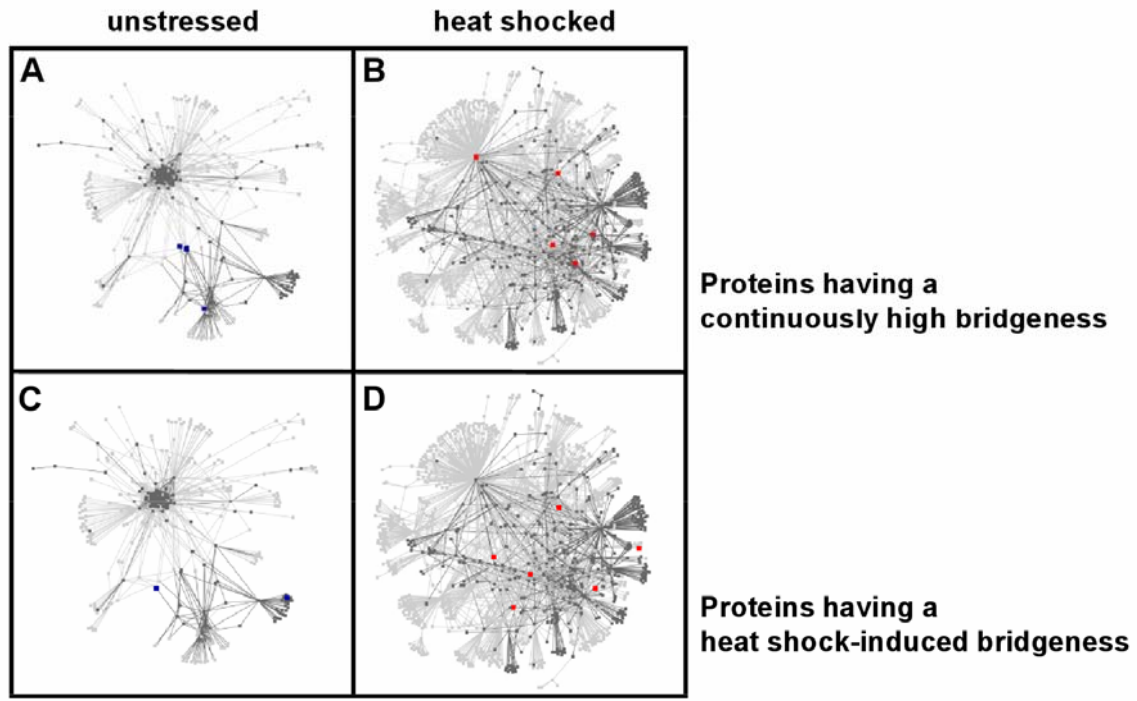

**Figure S6. Topological positions of proteins with persistent, or heat shock-induced bridgeness in the network of the strongest interactions of the yeast interactome.** Protein-protein interaction networks of unstressed (panels A and C) and heat shocked (15 min heat shock at 37°C; panels B and D) yeast cells were created as described in Methods of the main text. The subnetworks of their strongest links were determined and visualized as described in the legend of Figure 1 in the main text. Light grey colors denote the top 4%, while dark-grey colors the top 1% of interactions, respectively. Bridgeness of proteins was determined as described earlier [7]. Groups of proteins with persistent or heat shock-induced bridgeness (see Figure S6 and Table S4 of Text S1) are marked with larger blue filled circles in the unstressed conditions (panels A and C) and with larger red filled circles in the heat shocked conditions (panels B and D), respectively. Panels A and B. Topological positions of proteins having a persistently high bridgeness. Out of the 7 such proteins 3 and 5 proteins were visible in the subnetworks of the strongest links, before and after heat shock, respectively (see Figure 6 of the main text and Table S4 of Text S1). Panels C and D. Topological positions of proteins having a heat shock-induced bridgeness. Out of the 9 such proteins 2 and 6 proteins were visible in the subnetworks of the strongest links, before and after heat shock, respectively (see Figure 6 of the main text and Table S4 of Text S1). Bridges appeared at a larger ratio (31% compared to 69% before and after heat shock, respectively), and were re-organized to more inter-modular positions in the interactome of the strongest links after heat shock.

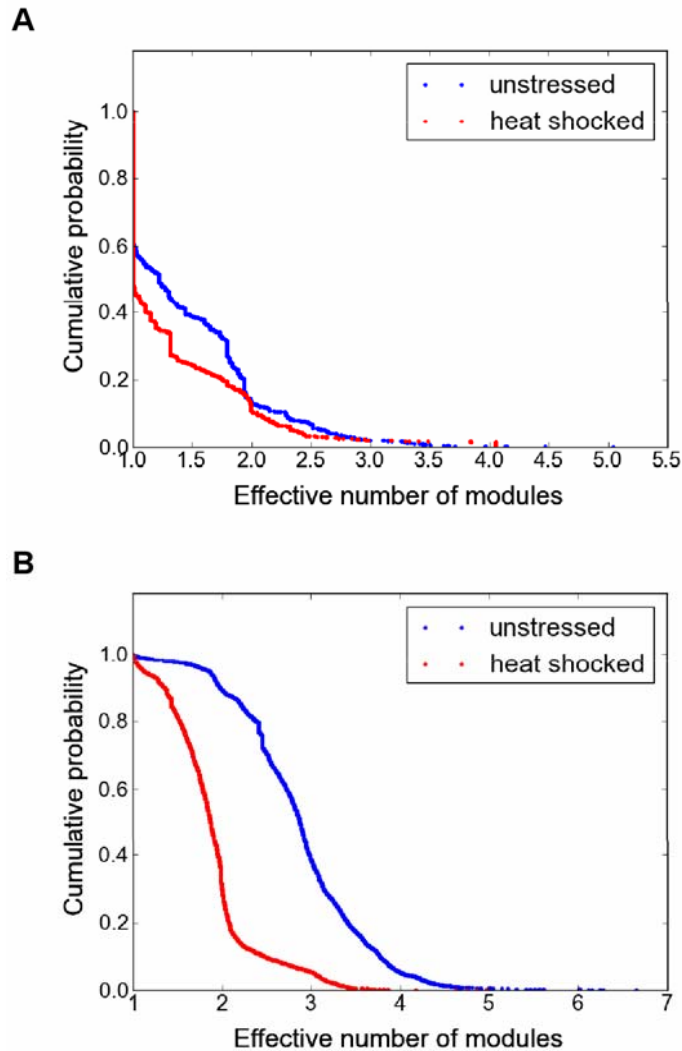

**Figure S7. Decrease of modular overlap of the yeast interactome using different model parameters.** Unstressed (blue dots) and heat shocked (15 min heat shock at 37°C, red dots) yeast protein-protein interaction networks using different model parameters were created as described in the Methods section of the main text. Overlapping modules were calculated by the NodeLand influence function method combined with the ProportionalHill module membership assignment method [7] as described in Methods of the main text. Panel A. Modular overlap of the high-confidence PPI dataset of Ekman et al. [8] in unstressed condition and upon heat shock. Here multiplication of the two node's mRNA abundances as link weights was used. The overlap of yeast interaction modules was represented by the cumulative distribution of the effective number of modules of yeast proteins (see Methods of the main text for more details). Upon heat shock the effective number of modules, that a yeast protein simultaneously belongs to, was significantly decreased (Wilcoxon paired test, $p < 2.2*10^{-16}$). Panel B. Modular overlap of the yeast BioGRID interactome in unstressed condition and upon heat shock, where mRNA expression data were logarithmically transformed. Here relative changes in mRNA levels were added to the baseline abundances and averaging of the two node's mRNA abundances as link weights was used (since logarithm transforms multiplication to addition). Upon heat shock the effective number of modules, that a yeast protein simultaneously belongs to, was significantly decreased (Wilcoxon paired test, $p < 2.2*10^{-16}$).

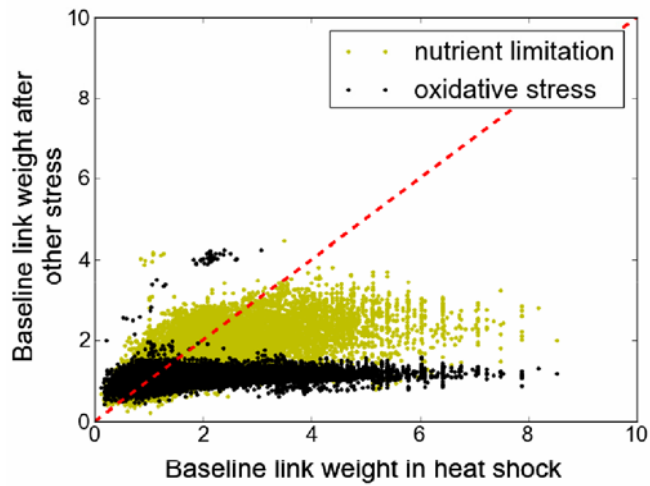

**Figure S8. Variability of interaction weights of unstressed yeast interactomes derived from relative changes of mRNA expression in different stress conditions.** Interaction weights of the yeast BioGRID protein-protein interactions networks (see the legend of Figure S1 of Text S1 and Methods of the main text for more details) were generated by using relative changes of mRNA expressions upon heat shock (15 min heat shock at 37°C), nutrient limitation (0.5 h amino acid starvation) and oxidative stress (20 min menadione). Weights were generated using a model, where relative changes upon stress were split to two changes: one-half for baseline weights and the other half for stressed weights (for more details see Methods section of the main text). Baseline interaction weights generated from changes upon heat shock were used as a reference, and the respective weights of nutrient limitation (yellow dots) and oxidative stress (black dots) were plotted as the function of the heat shock-related baseline interaction weights. The dashed red line represents a full correlation. The figure shows that oxidative stress- and nutrient limitation-related link weights are generally smaller than those in heat shock, which is due to the generally lower changes in mRNA levels in these two types of stresses than in heat shock.

# Supporting Tables

**Table S1. Functional annotation of selected yeast interactome modules at different thresholds of core proteins**

| GO ID | GO term: biological process | Number of core genes | Frequency of GO-term | p-value[a] | Threshold of core genes[b] |
|---|---|---|---|---|---|
| **A. An example of central interactome modules of unstressed yeast cells[c]** | | | | | |
| 6412 | translation | 5 genes | 5 out of 5 genes, 100.0% | $3.10 \times 10^{-4}$ | 79.20% |
| | | 25 genes | 25 out of 25 genes, 100.0% | $4.18 \times 10^{-24}$ | 56.87% |
| | | 50 genes | 50 out of 50 genes, 100.0% | $8.79 \times 10^{-50}$ | 42.22% |
| | | 75 genes | 73 out of 75 genes, 97.3% | $2.62 \times 10^{-70}$ | 53.41% |
| | | 100 genes | 81 out of 100 genes, 81.0% | $1.85 \times 10^{-62}$ | 0.20% |
| **B. An example of central interactome modules of heat shocked yeast cells** | | | | | |
| 6007 | glucose catabolic process | 5 genes | 5 out of 5 genes, 100.0% | $1.31 \times 10^{-10}$ | 49.01% |
| | | 25 genes | 8 out of 25 genes, 32.0% | $6.80 \times 10^{-11}$ | 4.56% |
| | | 50 genes | 10 out of 50 genes, 20.0% | $2.01 \times 10^{-11}$ | 0.27% |
| | | 75 genes | 13 out of 75 genes, 17.3% | $1.81 \times 10^{-14}$ | 0.12% |
| | | 100 genes | 13 out of 100 genes, 13.0% | $9.92 \times 10^{-13}$ | 0.07% |

[a] Statistical analysis was performed by hypergeometric test.
[b] Threshold is defined as the lowest community centrality value divided by the highest community centrality value (in per cent) of the respective module core.
[c] The examples show the preservation of the original GO-term at a statistically significant manner as a wider and wider selection of modular constituents are examined.

**Table S2. Functional annotation of yeast interactome modules in unstressed state and after heat shock**

| Module ranking[a] | Module centrality[b] | GO ID | GO term: biological process | p-value[c] | Top 5 gene names[d] |
|---|---|---|---|---|---|
| **A. Interactome modules of unstressed yeast cells** | | | | | |
| 1 | $7.16*10^6$ | 6412 | translation | $3.10*10^{-4}$ | YLR075W, YGL135W, YIL018W, YGR085C, YBL072C |
| 2 | $5.74*10^6$ | 6412 | translation | $1.80*10^{-4}$ | YBR191W, YGR085C, YGL103W, YNL178W, YBR031W |
| 3 | $1.72*10^6$ | 5975 | carbohydrate metabolic process | $4.71*10^{-6}$ | YKL060C, YCR012W, YHR174W, YGR192C, YOL086C |
| 4 | $2.13*10^5$ | 51276 | chromosome organization | $4.25*10^{-5}$ | YBR010W, YNL031C, YDR225W, YNL030W, YBL003C |
| 5 | $9.62*10^4$ | 51453 | regulation of intracellular pH | $4.89*10^{-3}$ | YBR106W, YHR026W, YGR060W, YLR372W, YEL027W |
| 6 | $1.08*10^4$ | - | no significant biological process was found | $<1.00*10^{-2}$ | YER117W, YBL187C, YER148W, YFR034C, YNL209W |
| 7 | $4.54*10^3$ | 7010 | cytoskeleton organization[e] | $1.90*10^{-6}$ | YFL039C, YLL050C, YHR179W, YBR109C, YLR291C[f] |
| 8 | $4.35*10^3$ | 43161 | proteasomal ubiquitin-dependent protein catabolic process | $1.97*10^{-3}$ | YDR328C, YHR021C, YLR167W, YDL126C, YMR276W |
| 9 | $4.09*10^3$ | 10499 | proteasomal ubiquitin-independent protein catabolic process | $4.96*10^{-13}$ | YER094C, YPR103W, YGR135W, YOL038W, YOR362C |
| 10 | $2.22*10^3$ | 48193 | Golgi vesicle transport | $7.62*10^{-5}$ | YDL192W, YDL137W, YPR110C, YKL196C, YDL226C |
| 11 | $1.70*10^3$ | 42254 | ribosome biogenesis | $3.88*10^{-5}$ | YKR057W, YGR285C, YER036C, YJL136C, YCL059C |

**Table S2. Functional annotation of yeast interactome modules in unstressed state and after heat shock** (continued)

| Module ranking[a] | Module centrality[b] | GO ID | GO term: biological process | p-value[c] | Top 5 gene names[d] |
|---|---|---|---|---|---|
| **A. Interactome modules of unstressed yeast cells** (continued) | | | | | |
| 12 | $8.88*10^2$ | 6360 | transcription from RNA polymerase I promoter[g] | $5.82*10^{-10}$ | YOR210W, YOR224C, YPR187W, YBR154C, YPR010C |
| 13 | $8.68*10^2$ | 7186 | G protein coupled receptor protein signaling pathway[e] | $1.05*10^{-3}$ | YGR037C, YFL026W, YLR359W, YBL079W, Q0110[h] |
| 14 | $8.20*10^2$ | 32543 | mitochondrial translation | $3.81*10^{-8}$ | YPL013C, YNL306W, YDR041W, YBR146W, YGR084C |
| 15 | $5.28*10^2$ | 51169 | nuclear transport | $4.90*10^{-7}$ | YDR002W, YLR293C, YMR235C, YGR218W, YKR048C |
| **B. Interactome modules of heat shocked yeast cells** | | | | | |
| 1 | $3.38*10^6$ | 10506 | regulation of autophagy | $9.00*10^{-4}$ | YDR343C, YDR342C, YLL039C, YKL203C, YJR066W |
| 2 | $2.88*10^6$ | 5975 | carbohydrate metabolic process | $4.71*10^{-6}$ | YCR012W, YKL060C, YGR192C, YHR174W, YOL086C |
| 3 | $4.15*10^5$ | 42026 | protein refolding | $9.00*10^{-4}$ | YLL026W, YOR027W, YLR216C, YOL013C, YOR244W |
| 4 | $2.37*10^5$ | 22402 | cell cycle process | $5.00*10^{-3}$ | YFL014W, YDR155C, YBR109C, YHR152W, YNL312W |
| 5 | $1.31*10^5$ | 6412 | translation | $3.10*10^{-4}$ | YDR382W, YLR075W, YGL103W, YDL083C, YOR063W |
| 6 | $5.55*10^4$ | 5992 | trehalose biosynthetic process | $1.80*10^{-4}$ | YBR126C, YMR251W, YCL040W, YML100W, YOL133W |
| 7 | $5.01*10^4$ | - | no significant biological process was found | $<1.00*10^{-2}$ | YNL160W, YBL032W, YOL133W, YDR134C, YNL055C |
| 8 | $3.55*10^4$ | 42026 | protein refolding[e] | $5.36*10^{-5}$ | YJR009C, YJL034W, YPR080W, YJL052W, YPL106C[i] |

**Table S2. Functional annotation of yeast interactome modules in unstressed state and after heat shock** (continued)

| Module ranking[a] | Module centrality[b] | GO ID | GO term: biological process | p-value[c] | Top 5 gene names[d] |
|---|---|---|---|---|---|
| **B. Interactome modules of heat shocked yeast cells** (continued) | | | | | |
| 9 | $3.45*10^4$ | 34284 | response to monosaccharide stimulus | $8.20*10^{-4}$ | YDR134C, YHR135C, YBL032W, YNL055C, YNL154C |
| 10 | $2.45*10^4$ | 48519 | negative regulation of biological process | $6.48*10^{-5}$ | YNL031C, YER177W, YBR010W, YDR099W, YOR244W |
| 11 | $3.49*10^3$ | 46467 | membrane lipid biosynthetic process | $2.56*10^{-3}$ | YDR276C, YDL212W, YBR036C, YBR183W, YDR307W |
| 12 | $2.36*10^3$ | 45454 | cell redox homeostasis | $1.48*10^{-3}$ | YLR109W, YGR209C, YIL035C, YJL141C, YMR059W |
| 13 | $1.10*10^3$ | 32543 | mitochondrial translation | $1.07*10^{-5}$ | YGR220C, YBL038W, YNL005C, YCR046C, YDR116C |
| 14 | $5.83*10^2$ | 55072 | iron homeostasis | $7.93*10^{-5}$ | YPL135W, YDR100W, YDL120W, YCL017C, YOL082W |

[a]Module ranking is based on the community centrality value of the most central protein of the respective module.
[b]Module centrality is defined as community centrality value of the most central protein of the respective module.
[c]Statistical analysis was performed by hypergeometric test.
[d]Gene ORF names are listed in decreasing order of their community centrality values.
[e]GO term annotation from 8 genes.
[f]YLR429W, YGL106W, YNL138W are the additional 3 genes for annotation.
[g]Transcription from RNA polymerase III (GO ID:6383, $p=8.95*10^{-8}$) and II promoter (GO ID:6366, $p=1.18*10^{-3}$) are also significant for these genes.
[h]YKL130W, YHR005C, YOR212W are the additional 3 genes for annotation.
[i]YMR186W, YOR136W, YPL240C are the additional 3 genes for annotation.

**Table S3. Yeast proteins with altered community centrality upon heat shock**

| ORF | Gene name | Functional annotation[a] |
|---|---|---|
| **A. small → large centrality[b] upon heat shock** | | |
| **YBL036C** | YBL036C | Putative non-specific single-domain racemase based on structural similarity |
| **YCL040W** | GLK1 | Glucokinase, catalyzes the phosphorylation of glucose at C6 in the first irreversible step of glucose metabolism |
| **YDR171W** | HSP42 | Small heat shock protein (sHSP) with chaperone activity |
| **YDR342C** | HXT7 | High-affinity glucose transporter of the major facilitator superfamily, nearly identical to Hxt6p, expressed at high basal levels relative to other HXTs, expression repressed by high glucose levels |
| **YDR343C** | HXT6 | High-affinity glucose transporter of the major facilitator superfamily, nearly identical to Hxt7p, expressed at high basal levels relative to other HXTs, repression of expression by high glucose requires SNF3 |
| **YER103W** | SSA4 | Heat shock protein that is highly induced upon stress |
| **YER125W** | RSP5 | E3 ubiquitin ligase of the NEDD4 family |
| **YFL014W** | HSP12 | Plasma membrane localized protein that protects membranes from desiccation |
| **YJR066W** | TOR1 | PIK-related protein kinase and rapamycin target |
| **YKL203C** | TOR2 | PIK-related protein kinase and rapamycin target |
| **YLL026W** | HSP104 | Heat shock protein that cooperates with Ydj1p (Hsp40) and Ssa1p (Hsp70) to refold and reactivate previously denatured, aggregated proteins |
| **YLR019W** | PSR2 | Functionally redundant Psr1p homolog, a plasma membrane phosphatase involved in the general stress response |
| **YOL100W** | PKH2 | Serine/threonine protein kinase involved in sphingolipid-mediated signaling pathway that controls endocytosis |
| **YOR181W** | LAS17 | Actin assembly factor, activates the Arp2/3 protein complex that nucleates branched actin filaments |
| **B. large community centrality[b] in both conditions** | | |
| **YCR012W** | PGK1 | 3-phosphoglycerate kinase, catalyzes transfer of high-energy phosphoryl groups from the acyl phosphate of 1,3-bisphosphoglycerate to ADP to produce ATP |
| **YDL229W** | SSB1 | Cytoplasmic ATPase that is a ribosome-associated molecular chaperone, functions with J-protein partner Zuo1p |
| **YDR050C** | TPI1 | Triose phosphate isomerase, abundant glycolytic enzyme |
| **YDR161W** | YDR161W | Putative protein of unknown function |
| **YDR188W** | CCT6 | Subunit of the cytosolic chaperonin CCT ring complex, related to Tcp1p, essential protein that is required for the assembly of actin and tubulins in vivo |
| **YDR510W** | SMT3 | Ubiquitin-like protein of the SUMO family, conjugated to lysine residues of target proteins |
| **YGR192C** | TDH3 | Glyceraldehyde-3-phosphate dehydrogenase, isozyme 3, involved in glycolysis and gluconeogenesis |
| **YGR252W** | GCN5 | Histone acetyltransferase, acetylates N-terminal lysines on histones H2B and H3 |
| **YHR174W** | ENO2 | Enolase II, a phosphopyruvate hydratase that catalyzes the conversion of 2-phosphoglycerate to phosphoenolpyruvate during glycolysis and the reverse reaction during gluconeogenesis |
| **YHR200W** | RPN10 | Non-ATPase base subunit of the 19S regulatory particle (RP) of the 26S proteasome |
| **YKL060C** | FBA1 | Fructose 1,6-bisphosphate aldolase, required for glycolysis and gluconeogenesis |
| **YKL152C** | GPM1 | Tetrameric phosphoglycerate mutase, mediates the conversion of 3-phosphoglycerate to 2-phosphoglycerate during glycolysis and the reverse reaction during gluconeogenesis |
| **YLL039C** | UBI4 | Ubiquitin, becomes conjugated to proteins, marking them for selective degradation via the ubiquitin-26S proteasome system |

**Table S3. Yeast proteins with altered community centrality upon heat shock** (continued)

| ORF | Gene name | Functional annotation[a] |
|---|---|---|
| **B. large community centrality[b] in both conditions** (continued) ||| 
| **YLR044C** | PDC1 | Major of three pyruvate decarboxylase isozymes, key enzyme in alcoholic fermentation, decarboxylates pyruvate to acetaldehyde |
| **YNL127W** | FAR11 | Protein involved in recovery from cell cycle arrest in response to pheromone, in a Far1p-independent pathway |
| **YOL086C** | ADH1 | Alcohol dehydrogenase, fermentative isozyme active as homo- or hetero-tetramers |
| **YOR098C** | NUP1 | Nuclear pore complex (NPC) subunit, involved in protein import/export and in export of RNAs, possible karyopherin release factor that accelerates release of karyopherin-cargo complexes after transport across NPC |
| **C. extra large → slightly smaller community centrality[b] upon heat shock** |||
| **YBL027W** | RPL19B | Protein component of the large (60S) ribosomal subunit, nearly identical to Rpl19Ap and has similarity to rat L19 ribosomal protein |
| **YBL072C** | RPS8A | Protein component of the small (40S) ribosomal subunit |
| **YBL092W** | RPL32 | Protein component of the large (60S) ribosomal subunit, has similarity to rat L32 ribosomal protein |
| **YBR031W** | RPL4A | N-terminally acetylated protein component of the large (60S) ribosomal subunit, nearly identical to Rpl4Bp and has similarity to E. coli L4 and rat L4 ribosomal proteins |
| **YBR048W** | RPS11B | Protein component of the small (40S) ribosomal subunit |
| **YBR181C** | RPS6B | Protein component of the small (40S) ribosomal subunit |
| **YBR189W** | RPS9B | Protein component of the small (40S) ribosomal subunit |
| **YBR191W** | RPL21A | Protein component of the large (60S) ribosomal subunit, nearly identical to Rpl21Bp and has similarity to rat L21 ribosomal protein |
| **YCR031C** | RPS14A | Ribosomal protein 59 of the small subunit, required for ribosome assembly and 20S pre-rRNA processing |
| **YDL075W** | RPL31A | Protein component of the large (60S) ribosomal subunit, nearly identical to Rpl31Bp and has similarity to rat L31 ribosomal protein |
| **YDL082W** | RPL13A | Protein component of the large (60S) ribosomal subunit, nearly identical to Rpl13Bp |
| **YDL083C** | RPS16B | Protein component of the small (40S) ribosomal subunit |
| **YDL136W** | RPL35B | Protein component of the large (60S) ribosomal subunit, identical to Rpl35Ap and has similarity to rat L35 ribosomal protein |
| **YDR012W** | RPL4B | Protein component of the large (60S) ribosomal subunit, nearly identical to Rpl4Ap and has similarity to E. coli L4 and rat L4 ribosomal proteins |
| **YDR025W** | RPS11A | Protein component of the small (40S) ribosomal subunit |
| **YDR064W** | RPS13 | Protein component of the small (40S) ribosomal subunit |
| **YDR382W** | RPP2B | Ribosomal protein P2 beta, a component of the ribosomal stalk, which is involved in the interaction between translational elongation factors and the ribosome |
| **YDR418W** | RPL12B | Protein component of the large (60S) ribosomal subunit, nearly identical to Rpl12Ap |
| **YDR447C** | RPS17B | Ribosomal protein 51 (rp51) of the small (40s) subunit |
| **YDR471W** | RPL27B | Protein component of the large (60S) ribosomal subunit, nearly identical to Rpl27Ap and has similarity to rat L27 ribosomal protein |
| **YER074W** | RPS24A | Protein component of the small (40S) ribosomal subunit |
| **YER102W** | RPS8B | Protein component of the small (40S) ribosomal subunit |
| **YFR031C-A** | RPL2A | Protein component of the large (60S) ribosomal subunit, identical to Rpl2Bp and has similarity to E. coli L2 and rat L8 ribosomal proteins |
| **YGL030W** | RPL30 | Protein component of the large (60S) ribosomal subunit, has similarity to rat L30 ribosomal protein |
| **YGL031C** | RPL24A | Ribosomal protein L30 of the large (60S) ribosomal subunit, nearly identical to Rpl24Bp and has similarity to rat L24 ribosomal protein |

**Table S3. Yeast proteins with altered community centrality upon heat shock** (continued)

| ORF | Gene name | Functional annotation[a] |
|---|---|---|
| **C. extra large → slightly smaller community centrality[b] upon heat shock** (continued) | | |
| **YGL076C** | RPL7A | Protein component of the large (60S) ribosomal subunit, nearly identical to Rpl7Bp and has similarity to E. coli L30 and rat L7 ribosomal proteins |
| **YGL103W** | RPL28 | Ribosomal protein of the large (60S) ribosomal subunit, has similarity to E. coli L15 and rat L27a ribosomal proteins |
| **YGL135W** | RPL1B | N-terminally acetylated protein component of the large (60S) ribosomal subunit, nearly identical to Rpl1Ap and has similarity to E. coli L1 and rat L10a ribosomal proteins |
| **YGL147C** | RPL9A | Protein component of the large (60S) ribosomal subunit, nearly identical to Rpl9Bp and has similarity to E. coli L6 and rat L9 ribosomal proteins |
| **YGR034W** | RPL26B | Protein component of the large (60S) ribosomal subunit, nearly identical to Rpl26Ap and has similarity to E. coli L24 and rat L26 ribosomal proteins |
| **YGR085C** | RPL11B | Protein component of the large (60S) ribosomal subunit, nearly identical to Rpl11Ap |
| **YHL001W** | RPL14B | Protein component of the large (60S) ribosomal subunit, nearly identical to Rpl14Ap and has similarity to rat L14 ribosomal protein |
| **YHL033C** | RPL8A | Ribosomal protein L4 of the large (60S) ribosomal subunit, nearly identical to Rpl8Bp and has similarity to rat L7a ribosomal protein |
| **YHR141C** | RPL42B | Protein component of the large (60S) ribosomal subunit, identical to Rpl42Ap and has similarity to rat L44 |
| **YHR203C** | RPS4B | Protein component of the small (40S) ribosomal subunit |
| **YIL018W** | RPL2B | Protein component of the large (60S) ribosomal subunit, identical to Rpl2Ap and has similarity to E. coli L2 and rat L8 ribosomal proteins |
| **YIL069C** | RPS24B | Protein component of the small (40S) ribosomal subunit |
| **YIL133C** | RPL16A | N-terminally acetylated protein component of the large (60S) ribosomal subunit, binds to 5.8 S rRNA |
| **YJL177W** | RPL17B | Protein component of the large (60S) ribosomal subunit, nearly identical to Rpl17Ap and has similarity to E. coli L22 and rat L17 ribosomal proteins |
| **YJL190C** | RPS22A | Protein component of the small (40S) ribosomal subunit |
| **YJR123W** | RPS5 | Protein component of the small (40S) ribosomal subunit, the least basic of the non-acidic ribosomal proteins |
| **YJR145C** | RPS4A | Protein component of the small (40S) ribosomal subunit |
| **YKL180W** | RPL17A | Protein component of the large (60S) ribosomal subunit, nearly identical to Rpl17Bp and has similarity to E. coli L22 and rat L17 ribosomal proteins |
| **YLL045C** | RPL8B | Ribosomal protein L4 of the large (60S) ribosomal subunit, nearly identical to Rpl8Ap and has similarity to rat L7a ribosomal protein |
| **YLR029C** | RPL15A | Protein component of the large (60S) ribosomal subunit, nearly identical to Rpl15Bp and has similarity to rat L15 ribosomal protein |
| **YLR075W** | RPL10 | Protein component of the large (60S) ribosomal subunit, responsible for joining the 40S and 60S subunits |
| **YLR340W** | RPP0 | Conserved ribosomal protein P0 of the ribosomal stalk, which is involved in interaction between translational elongation factors and the ribosome |
| **YLR441C** | RPS1A | Ribosomal protein 10 (rp10) of the small (40S) subunit |
| **YLR448W** | RPL6B | Protein component of the large (60S) ribosomal subunit, has similarity to Rpl6Ap and to rat L6 ribosomal protein |
| **YML063W** | RPS1B | Ribosomal protein 10 (rp10) of the small (40S) subunit |
| **YML073C** | RPL6A | N-terminally acetylated protein component of the large (60S) ribosomal subunit, has similarity to Rpl6Bp and to rat L6 ribosomal protein |
| **YMR194W** | RPL36A | N-terminally acetylated protein component of the large (60S) ribosomal subunit, nearly identical to Rpl36Bp and has similarity to rat L36 ribosomal protein |

**Table S3. Yeast proteins with altered community centrality upon heat shock** (continued)

| ORF | Gene name | Functional annotation[a] |
|---|---|---|
| **C. extra large → slightly smaller community centrality[b] upon heat shock** (continued) | | |
| **YMR242C** | RPL20A | Protein component of the large (60S) ribosomal subunit, nearly identical to Rpl20Bp and has similarity to rat L18a ribosomal protein |
| **YNL069C** | RPL16B | N-terminally acetylated protein component of the large (60S) ribosomal subunit, binds to 5.8 S rRNA |
| **YNL096C** | RPS7B | Protein component of the small (40S) ribosomal subunit, nearly identical to Rps7Ap |
| **YNL178W** | RPS3 | Protein component of the small (40S) ribosomal subunit, has apurinic/apyrimidinic (AP) endonuclease activity |
| **YNL301C** | RPL18B | Protein component of the large (60S) ribosomal subunit, identical to Rpl18Ap and has similarity to rat L18 ribosomal protein |
| **YOL040C** | RPS15 | Protein component of the small (40S) ribosomal subunit |
| **YOL127W** | RPL25 | Primary rRNA-binding ribosomal protein component of the large (60S) ribosomal subunit, has similarity to E. coli L23 and rat L23a ribosomal proteins |
| **YOR063W** | RPL3 | Protein component of the large (60S) ribosomal subunit, has similarity to E. coli L3 and rat L3 ribosomal proteins |
| **YOR096W** | RPS7A | Protein component of the small (40S) ribosomal subunit, nearly identical to Rps7Bp |
| **YOR312C** | RPL20B | Protein component of the large (60S) ribosomal subunit, nearly identical to Rpl20Ap and has similarity to rat L18a ribosomal protein |
| **YPL090C** | RPS6A | Protein component of the small (40S) ribosomal subunit |
| **YPL131W** | RPL5 | Protein component of the large (60S) ribosomal subunit with similarity to E. coli L18 and rat L5 ribosomal proteins |
| **YPL198W** | RPL7B | Protein component of the large (60S) ribosomal subunit, nearly identical to Rpl7Ap and has similarity to E. coli L30 and rat L7 ribosomal proteins |
| **YPL220W** | RPL1A | N-terminally acetylated protein component of the large (60S) ribosomal subunit, nearly identical to Rpl1Bp and has similarity to E. coli L1 and rat L10a ribosomal proteins |
| **D. large → small community centrality[b] upon heat shock** | | |
| **YDL014W** | NOP1 | Nucleolar protein, component of the small subunit processome complex, which is required for processing of pre-18S rRNA |
| **YDR450W** | RPS18A | Protein component of the small (40S) ribosomal subunit |
| **YGR118W** | RPS23A | Ribosomal protein 28 (rp28) of the small (40S) ribosomal subunit, required for translational accuracy |
| **YHR010W** | RPL27A | Protein component of the large (60S) ribosomal subunit, nearly identical to Rpl27Bp and has similarity to rat L27 ribosomal protein |
| **YLR287C-A** | RPS30A | Protein component of the small (40S) ribosomal subunit |
| **YNL110C** | NOP15 | Constituent of 66S pre-ribosomal particles, involved in 60S ribosomal subunit biogenesis |
| **YNL124W** | NAF1 | RNA-binding protein required for the assembly of box H/ACA snoRNPs and thus for pre-rRNA processing, forms a complex with Shq1p and interacts with H/ACA snoRNP components Nhp2p and Cbf5p |
| **YNL132W** | KRE33 | Essential protein, required for biogenesis of the small ribosomal subunit |
| **E. small community centrality[b] in both conditions** | | |
| **YDR104C** | SPO71 | Meiosis-specific protein of unknown function, required for spore wall formation during sporulation |
| **YDR222W** | YDR222W | Protein of unknown function |
| **YEL072W** | RMD6 | Protein required for sporulation |

**Table S3. Yeast proteins with altered community centrality upon heat shock** (continued)

| ORF | Gene name | Functional annotation[a] |
|---|---|---|
| **E. small community centrality[b] in both conditions** (continued) | | |
| **YFL056C** | AAD6 | Putative aryl-alcohol dehydrogenase with similarity to *P. chrysosporium* aryl-alcohol dehydrogenase, involved in the oxidative stress response |
| **YGL263W** | COS12 | Protein of unknown function, member of the DUP380 subfamily of conserved, often subtelomerically-encoded proteins |
| **YJR055W** | HIT1 | Protein of unknown function, required for growth at high temperature |
| **YJR129C** | YJR129C | Putative protein of unknown function |
| **YLL052C** | AQY2 | Water channel that mediates the transport of water across cell membranes, only expressed in proliferating cells, controlled by osmotic signals, may be involved in freeze tolerance |
| **YLR010C** | TEN1 | Protein that regulates telomeric length |
| **YLR213C** | CRR1 | Putative glycoside hydrolase of the spore wall envelope |

[a]Functional annotation of yeast proteins was achieved by GO annotation database downloaded from http://www.geneontology.org/GO.downloads.annotations.shtml at 10/9/2010.
[b]Community centrality values of proteins were calculated by the NodeLand influence function method [7]. Groups are the same as on Figure 4 of the main text.

**Table S4. Yeast proteins with altered bridgeness upon heat shock**

| ORF | Gene name | Functional annotation[a] |
|---|---|---|
| **The 4 most induced bridges of the 9 heat-induced bridges**[b] | | |
| **YNL103W** | MET4 | Leucine-zipper transcriptional activator, responsible for the regulation of the sulfur amino acid pathway, requires different combinations of the auxiliary factors Cbf1p, Met28p, Met31p and Met32p |
| **YNL189W**\*\* | SRP1 | Karyopherin alpha homolog, forms a dimer with karyopherin beta Kap95p to mediate import of nuclear proteins, binds the nuclear localization signal of the substrate during import |
| **YNL197C**\* | WHI3 | RNA binding protein that sequesters CLN3 mRNA in cytoplasmic foci |
| **YOL062C** | APM4 | Mu2-like subunit of the clathrin associated protein complex (AP-2) |
| **The 5 less induced bridges of the 9 heat-induced bridges**[b] | | |
| **YER021W**\* | RPN3 | Essential, non-ATPase regulatory subunit of the 26S proteasome lid, similar to the p58 subunit of the human 26S proteasome |
| **YER125W**\* | RSP5 | E3 ubiquitin ligase of the NEDD4 family |
| **YLR249W**\*\* | YEF3 | Gamma subunit of translational elongation factor eEF1B, stimulates the binding of aminoacyl-tRNA (AA-tRNA) to ribosomes by releasing eEF1A (Tef1p/Tef2p) from the ribosomal complex |
| **YNL113W** | RPC19 | RNA polymerase subunit AC19, common to RNA polymerases I and III |
| **YNL161W**\* | CBK1 | Serine/threonine protein kinase that regulates cell morphogenesis pathways |
| **7 persistent bridges**[b] | | |
| **YAR027W** | UIP3 | Putative integral membrane protein of unknown function |
| **YBL032W**\* | HEK2 | RNA binding protein involved in the asymmetric localization of ASH1 mRNA |
| **YDR510W**\*\* | SMT3 | Ubiquitin-like protein of the SUMO family, conjugated to lysine residues of target proteins |
| **YJL092W**\* | SRS2 | DNA helicase and DNA-dependent ATPase involved in DNA repair, needed for proper timing of commitment to meiotic recombination and transition from Meiosis I to II |
| **YLL039C**\*\* | UBI4 | Ubiquitin, becomes conjugated to proteins, marking them for selective degradation via the ubiquitin-26S proteasome system |
| **YMR236W** | TAF9 | Subunit (17 kDa) of TFIID and SAGA complexes, involved in RNA polymerase II transcription initiation and in chromatin modification, similar to histone H3 |
| **YOL135C**\*\* | MED7 | Subunit of the RNA polymerase II mediator complex |
| **18 persistent, albeit less dominant bridges**[b] | | |
| **YAL024C** | LTE1 | Protein similar to GDP/GTP exchange factors but without detectable GEF activity |
| **YBR009C** | HHF1 | Histone H4, core histone protein required for chromatin assembly and chromosome function |
| **YBR017C** | KAP104 | Transportin or cytosolic karyopherin beta 2 |
| **YCL028W** | RNQ1 | [PIN(+)] prion, an infectious protein conformation that is generally an ordered protein aggregate |
| **YDR188W** | CCT6 | Subunit of the cytosolic chaperonin Cct ring complex, related to Tcp1p, essential protein that is required for the assembly of actin and tubulins in vivo |
| **YDR347W** | MRP1 | Mitochondrial ribosomal protein of the small subunit |
| **YFR034C** | PHO4 | Basic helix-loop-helix (bHLH) transcription factor of the myc-family |
| **YGR124W** | ASN2 | Asparagine synthetase, isozyme of Asn1p |
| **YHR062C** | RPP1 | Subunit of both RNase MRP, which cleaves pre-rRNA, and nuclear RNase P, which cleaves tRNA precursors to generate mature 5' ends |
| **YHR152W** | SPO12 | Nucleolar protein of unknown function, positive regulator of mitotic exit |
| **YJR121W** | ATP2 | Beta subunit of the F1 sector of mitochondrial F1F0 ATP synthase, which is a large, evolutionarily conserved enzyme complex required for ATP synthesis |
| **YKL203C** | TOR2 | PIK-related protein kinase and rapamycin target |

**Table S4. Yeast proteins with altered bridgeness upon heat shock** (continued)

| ORF | Gene name | Functional annotation[a] |
|---|---|---|
| **18 persistent, albeit less dominant bridges**[b] **(continued)** | | |
| **YLR150W** | STM1 | Protein required for optimal translation under nutrient stress |
| **YLR340W** | RPP0 | Conserved ribosomal protein P0 of the ribosomal stalk, which is involved in interaction between translational elongation factors and the ribosome |
| **YMR024W** | MRPL3 | Mitochondrial ribosomal protein of the large subunit |
| **YOL027C** | MDM38 | Mitochondrial inner membrane protein, involved in membrane integration of a subset of mitochondrial proteins |
| **YOR098C** | NUP1 | Nuclear pore complex (NPC) subunit, involved in protein import/export and in export of RNAs, possible karyopherin release factor that accelerates release of karyopherin-cargo complexes after transport across NPC |
| **YPL178W** | CBC2 | Small subunit of the heterodimeric cap binding complex that also contains Sto1p, component of the spliceosomal commitment complex |
| **The most decreased bridge of the 3 heat-decreased bridges**[b] | | |
| **YNL110C** | NOP15 | Constituent of 66S pre-ribosomal particles, involved in 60S ribosomal subunit biogenesis |
| **The 2 less decreased bridges of the 3 heat-decreased bridges**[b] | | |
| **YDR172W** | SUP35 | Translation termination factor eRF3 |
| **YDR405W** | MRP20 | Mitochondrial ribosomal protein of the large subunit |

[a]Functional annotation of yeast proteins was achieved by GO annotation database downloaded from http://www.geneontology.org/GO.downloads.annotations.shtml at 10/9/2010.
[b]Bridgeness values of proteins were calculated as described earlier [7]. Groups are the same as on Figure 6 of the main text.
*Bridge appears as a node of the strongest links of the yeast interactome shown on Figure S6 in Text S1 in heat shock only.
** Bridge appears as a node of the strongest links of the yeast interactome shown on Figure S6 in Text S1 both in unstressed state and after heat shock.

# References

1. Stark C, Breitkreutz BJ, Chatr-Aryamontri A, Boucher L, Oughtred R, et al. (2011) The BioGRID interaction database: 2011 update. Nucleic Acids Res 39: D698–704.
2. Holstege FC, Jennings EG, Wyrick JJ, Lee TI, Hengartner CJ, et al. (1998) Dissecting the regulatory circuitry of a eukaryotic genome. Cell 95: 717–728.
3. Gasch AP, Spellman PT, Kao CM, Carmel-Harel O, Eisen MB, et al. (2000) Genomic expression programs in the response of yeast cells to environmental changes. Mol Biol Cell 11: 4241–4257.
4. Thomas GH, Zucker J, Macdonald SJ, Sorokin A, Goryanin I, et al. (2009) A fragile metabolic network adapted for cooperation in the symbiotic bacterium Buchnera aphidicola. BMC Syst Biol 3: 24.
5. Feist AM, Henry CS, Reed JL, Krummenacker M, Joyce AR, et al. (2007) A genome-scale metabolic reconstruction for Escherichia coli K-12 MG1655 that accounts for 1260 ORFs and thermodynamic information. Mol Syst Biol 3: 121.
6. Shannon P, Markiel A, Ozier O, Baliga NS, Wang JT, et al. (2003) Cytoscape: a software environment for integrated models of biomolecular interaction networks. Genome Res 13: 2498–2504.
7. Kovacs IA, Palotai R, Szalay MS, Csermely P (2010) Community landscapes: an integrative approach to determine overlapping network module hierarchy, identify key nodes and predict network dynamics. PLoS ONE 5: e12528.
8. Ekman D, Light S, Bjorklund AK, Elofsson A (2006) What properties characterize the hub proteins of the protein-protein interaction network of *Saccharomyces cerevisiae*? Genome Biol 7: R45.